\documentclass[twocolumn]{aastex63}

\shorttitle{XCN and aromatic C\sbond D features in the YSO}
\shortauthors{Onaka et al.}

\begin{document}

\title{Near-infrared spectroscopy of a massive young stellar object in the direction toward the Galactic Center:
XCN and aromatic C\sbond D features}

\correspondingauthor{Takashi Onaka}
\email{onaka@astron.s.u-tokyo.ac.jp}

\author[0000-0002-8234-6747]{Takashi Onaka}
\affiliation{Department of Physics, Faculty of Science and Engineering, 
Meisei University, 2-1-1 Hodokubo, Hino, Tokyo 191-8506, Japan\\
}
\affiliation{Department of Astronomy, Graduate School of Science, The University of Tokyo, 7-3-1 Hongo, Bunkyo-ku, Tokyo 113-0033, Japan\\
}

\author[0000-0001-7641-5497]{Itsuki Sakon}
\affiliation{Department of Astronomy, Graduate School of Science, The University of Tokyo, 7-3-1 Hongo, Bunkyo-ku, Tokyo 113-0033, Japan\\
}
\author[0000-0002-0095-3624]{Takashi Shimonishi}
\affiliation{Institute of Science and Technology, Niigata University, Ikarashi-nihoncho 8050, Nishi-ku, Niigata 950-2181, Japan}



\begin{abstract}
We report near-infrared (2.5--5\,$\mu$m) long-slit ($\sim 30\arcsec$) spectroscopy of a young stellar object in the direction toward the Galactic center with
the Infrared Camera on board the AKARI satellite.  The present target is suggested to be AFGL\,2006 based on its very red color and close location.  
The spectra show
strong absorption features of H$_2$O and CO$_2$ ices, and emission of \ion{H}{1} Br$\alpha$ recombination line and the 3.3\,$\mu$m band, 
the latter of which originates from
polycyclic aromatic hydrocarbons (PAHs) or materials containing PAHs.  The spectra show a 
broad, complex absorption feature at 4.65\,$\mu$m, which is well explained by a combination of absorption features of CO ice, CO gas, and XCN, and \ion{H}{1} Pf$\beta$ emission.
The spectra also indicate excess emission at 4.4\,$\mu$m.  
The characteristics of the spectra suggest that the object is a massive young stellar object.
The XCN feature shows a good 
correlation with the Br$\alpha$ emission, suggesting that the photolysis by ultraviolet photons plays an important role in the formation of the XCN carriers, part of which are attributed to OCN$^-$.  
The 4.4\,$\mu$m emission shows a good correlation with the 3.3\,$\mu$m PAH emission, providing supporting evidence
that it comes from the aromatic C\sbond D stretching vibration.  The formation of OCN$^-$ is of importance for the formation process of prebiotic matter 
in the interstellar medium (ISM), while the detection of aromatic C\sbond D emission provides valuable information on the deuteration process of PAHs in the ISM and
implications on the hiding site of the missing deuterium in the ISM.
\end{abstract}



\section{Introduction} \label{sec:intro}

The near-infrared (NIR) spectrum of the interstellar medium (ISM) contains a number of gas lines and ice bands,
providing us with a wealth of information on the physical conditions of the gas and dust species in the object.  In the NIR, 
\ion{H}{2} region--photodissociation region (PDR) complexes emit a plethora of gas lines, including hydrogen and helium recombination lines, and
H$_2$ rovibrational transitions,
together with the 3.3--3.5\,$\mu$m features from aromatic and aliphatic C\sbond H bonds in polycyclic aromatic hydrocarbons (PAHs) or PAH containing
materials, while they sometimes show absorption features of 
H$_2$O and CO$_2$ ices at 3.0 and 4.26\,$\mu$m, respectively \citep[e.g.,][]{2014ApJ...780..114O, 2014ApJ...784...53M}.  PAHs are also thought to be the carriers of a series of the emission bands
observed in the mid-infrared \citep{2008ARA&A..46..289T,  2020NatAs...4..339L}.  Young stellar objects (YSOs) in the early evolutionary stage further show
absorption features of CO and "XCN" ices together with CO gas absorption in 4.6--4.9\,$\mu$m \citep[e.g.,][]{1984ApJ...276..533L, 2003A&A...408..981P, 2004ApJS..151...35G, 2021ApJ...916...75O}.
The XCN feature consists of two
components with peaks at 4.60 and  4.62\,$\mu$m.  The latter is securely ascribed to OCN$^-$ \citep{2005A&A...441..249V}.  While  CO absorbed 
at bare grain surfaces has been proposed \citep{2005MNRAS.356.1283F}, the carriers for the former component are not clearly identified \citep{2015ARA&A..53..541B}.
\citet{2011ApJ...740..109O} suggest that OCN$^-$ is responsible for the entire XCN feature based on the correlations with other ice species among YSOs.
The formation process of OCN$^{-}$ has been studied in various laboratory experiments, suggesting that there are several pathways to form OCN$^-$, but 
its actual formation process in the ISM is yet to be fully understood \citep[e.g.,][references therein]{1999ApJ...513..294P, 2004A&A...415..425V, 2010ApJ...723..641B, 2016MNRAS.460.4297F}.
The formation of OCN$^-$ is of importance for the understanding of the formation of prebiotic matter in the ISM \citep[e.g.,][]{2015MNRAS.446..439F}, 
and it may also be related to the formation of nitrogen-bearing PAHs recently discovered in dense clouds \citep{2018Sci...359..202M, 2021NatAs...5..181B, 2021Sci...371.1265M}.

PAHs are expected to be deuterated via several processes in the ISM \citep{2001M&PS...36.1117S}, and deuterated PAHs are proposed as a likely hiding site of missing deuterium (D) in the ISM \citep{2006ApJ...647.1106L, 2006ASPC..348...58D}.
Aromatic and aliphatic C\sbond D stretching vibrations of deuterated PAHs are supposed to appear at 4.4--4.7\,$\mu$m \citep{2004ApJ...614..770H, 2021ApJ...917L..35A}.   
While excess emission at 4.6--4.8\,$\mu$m, which can be attributed to the aliphatic C\sbond D bond, has been detected in several sources, the aromatic C\sbond D feature at around 4.4\,$\mu$m
is relatively elusive and detected only in a few sources with a significant level to date \citep{2004ApJ...604..252P, 2014ApJ...780..114O, 2016A&A...586A..65D}.  
Firm detection of the aromatic C\sbond D stretching feature would support the attribution of the excess emission at 4.6--4.8\,$\mu$m to the aliphatic C\sbond D stretching feature \citep{2011EAS....46..399B}.

In this paper, we present the results of a long-slit NIR spectroscopy of an envelope of a YSO in the direction toward the Galactic center with the Infrared
Camera (IRC) on board the AKARI satellite \citep{2007PASJ...59S.401O}.  The spectra show deep H$_2$O and CO$_2$ ice absorption features.
In the H$_2$O ice absorption feature, the 3.3\,$\mu$m feature of aromatic C\sbond H is superimposed in emission.  \ion{H}{1} Br$\alpha$ and Br$\beta$ emissions at 4.052 and 2.636\,$\mu$m are also seen in the spectra.
In the 4.5--4.8\,$\mu$m region, the observed spectra show a broad, complex absorption feature at around 4.65\,$\mu$m, which can be attributed to a combination of absorption of XCN, CO ice, and CO gas, and
\ion{H}{1} Pf$\beta$ emission.
We make a decomposition of the feature and discuss the spatial variation of the XCN component with other observed features.  
The observed spectra also
show an excess emission at around 4.4\,$\mu$m, which can be attributed to aromatic C\sbond D bonds.  We study the spatial variation of the 4.4\,$\mu$m excess emission and discuss the deuteration processes of PAHs in the present object.

In Section~\ref{sec:obs}, the observations and data reduction are described.   The results and analysis are presented in Section~\ref{sec:results}, and 
the discussion is given in Section~\ref{sec:discussion}.  Section~\ref{sec:summary} summaries the results of the present paper.

\section{Observations and data reduction} \label{sec:obs}

The present observations were carried out on 2008 September 19
(observation ID: 5200149.1)
during the warm mission period as part of the director time for the purpose of the calibration of the instrument.
They were made in the slit spectroscopy mode of the IRC with the NIR grism (NG) and NIR prism (NP) at the narrow, long-slit Nh of a size of $3\arcsec \times 1\arcmin$, which
provided spectra of 2.5--4.9\,$\mu$m with a spectral resolution of about 0.02\,$\mu$m with NG \citep{2007PASJ...59S.411O}.
The observations of the same position were made with NG a year later to monitor the performance of the instrument on 2009 September 20 (observations ID: 5200984.1).  They showed very similar NG spectra,
but had larger noises due to the increase of hot pixels in the detector for the later phase of the warm mission period \citep{2010SPIE.7731E..0MO}. 
Therefore, we decided to use only the NG data of observations 5200149.1 for the present study. The NP data have lower spectral resolution and are not used.
The observations consisted of five frames with NG, four frames with NP, and one frame of imaging data with the N3 filter
(3.2\,$\mu$m).  The imaging data were taken for the purpose of the accurate position determination.  The net integration time of the NG observations was about 245\,s.

\begin{figure*}[ht!]
\epsscale{0.9}
\plotone{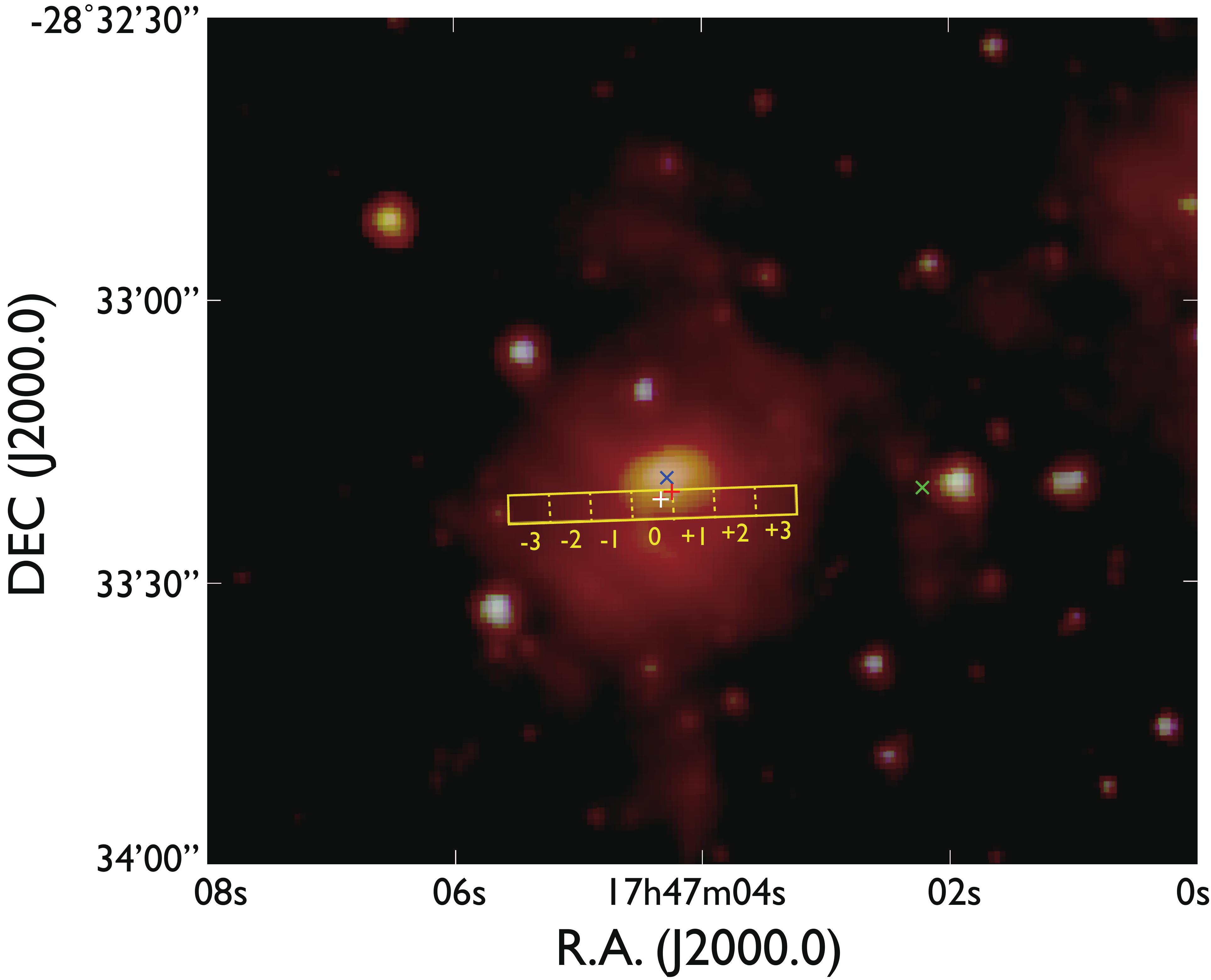}
\caption{Location of the Nh slit of the IRC on the IRAC three-color image of the target position (blue, IRAC band 1; green, IRAC band 2; and red, IRAC band 3).   The IRAC images are
taken from \citet{https://doi.org/10.26131/irsa543}.
The Nh slit of a $30\farcs$66 $\times 3\arcsec$ area is depicted by the solid yellow rectangle.
The numbers along the yellow rectangle indicate the position number of the extracted spectrum (see the text), increasing from $-3$ at the east end to $+3$ at the west end with an interval of $4\farcs$38. 
The boundaries of the spectrum extraction windows are indicated by the yellow dashed lines.  The slit center (position 0) is located at (R.A., decl. (J2000.0)) = (17:47:04.39, -28:33:21.8).   The green $\times$ marks the position of IRAS\,17438-2832,
while the blue $\times$ and the red $+$ indicate the positions of the Spitzer/IRAC source of SSTGC\,0753438 \citep{2008ApJS..175..147R} and 2MASS\,J17470424-2833204, respectively.  The while $+$ shows the position of the 
Hi-GAL source HIGALPB000.4901-0.0744 \citep{2016A&A...591A.149M}.
\label{fig:image}}
\end{figure*}

The data were reduced with the latest IRC spectroscopic toolkit for phase three (version 2018203)\footnote{ \url{https://www.ir.isas.jaxa.jp/AKARI/Observation/support/IRC/software/IRC_SPEC_TOOLKIT_P3_20181203.tar.gz}},
which included the latest wavelength calibration and the sensitivity correction due to the variation of the detector temperature \citep{2019PASJ...71....2B}.
Spectra of 2.5--4.9\,$\mu$m were extracted for every 4$\farcs$38 (3 pixels), which matched the point-spread function of the IRC in the warm mission \citep{2010SPIE.7731E..0MO}.  
We extracted seven spectra from the region, whose surface brightness is brighter than 50 and 70\,MJy\,sr$^{-1}$ at IRAC band 1 (3.6\,$\mu$m) and
band 2 (4.5\,$\mu$m), respectively. The center position of slit corresponds to the brightest part in the slit.  
The surface brightness decreases rapidly outside of the extracted region, and we assume that the emission within the region is mostly associated with the
central source.
The position of the slit was 
determined using the 3.2\,$\mu$m image, by matching the detected point sources with 
the Two Micron All Sky Survey (2MASS) catalog \citep{2006AJ....131.1163S}.  The imaging data were reduced with the IRC imaging toolkit for phase three
 (version 20131202)\footnote{ \url{https://www.ir.isas.jaxa.jp/AKARI/Observation/support/IRC/software/irc_p3_20131202.tar.gz}}.  The accuracy of the position determination was estimated to be
better than $1\arcsec$.  The slit position and extraction window for each spectrum are shown in Figure~\ref{fig:image} by the dashed lines in the yellow rectangle.
The extracted positions are denoted by $-3, -2,$ ..., and $+3$ from east to west as shown in Figure~\ref{fig:image}. 

\begin{deluxetable*}{ll}
\tablecaption{Position and infrared flux of cataloged point sources near the present target \label{tab:target}}
\tablewidth{0pt}
\tablehead{
\colhead{Source Name} &  \colhead{J2000 Coordinates}\\
\colhead{\quad Band Name (Wavelength)} & {\qquad Flux (Jy)}
}
\startdata
2MASS\,J17470424-2833204\tablenotemark{a}  & 17:47:04.24 \quad --28:33:20.4\\
\quad J (1.235 \,$\mu$m) & \quad $< 1.15 \times 10^{-4}$ \\
\quad H (1.662 \,$\mu$m) & \quad $(5.38 \pm 0.31) \times 10^{-3}$ \\
\quad K$_\mathrm{s}$ (2.159 \,$\mu$m) & \quad $ 0.0455 \pm 0.0025$\\
\hline
SSTGC\,0753438\tablenotemark{b}& 17:47:04.272 \quad --28:33:18.93 \\
\quad IRAC1 (3.6\,$\mu$m) & \quad $0.233 \pm 0.002$ \\
\quad IRAC2 (4.5\,$\mu$m) & \quad $0.511 \pm 0.003$ \\
\quad IRAC3 (5.8\,$\mu$m) & \quad $1.36 \pm 0.01$ \\
\quad IRAC4 (8.0\,$\mu$m) & \quad $3.38 \pm 0.02$ \\
\hline
WISE\,J174703.97-283317.0\tablenotemark{c} & 17:47:03.97 \quad --28:33:17.0\\
\quad W1 (3.4\,$\mu$m) &  \quad $0.246 \pm 0.006$ \\
\quad W2 (4.6\,$\mu$m) &  \quad $4.33 \pm 0.24$ \\
\hline
AKARI IRC\,1747040-283320\tablenotemark{d} & 17:47:04.0 \quad --28:33:20\\
\quad S9W (9\,$\mu$m) & \quad 26.5 \\
\quad L18W (18\,$\mu$m) & \quad 165.2 \\\hline
IRAS\,17438-2832 (AFGL\,2006)\tablenotemark{e} & 17:47:02.2 \quad  --28:33:20 \\
\quad 12\,$\mu$m & \quad 32.01 \\
\quad 25\,$\mu$m & \quad 245.1 \\
\quad 60\,$\mu$m & \quad $< 60.51$ \\
\quad 100\,$\mu$m & \quad 8659\\
\hline
HIGALPB000.4901-0.0744\tablenotemark{f} & 17:47:04.33  \quad --28:33:22.1 \\
\quad 70\,$\mu$m & \quad $284.603 \pm 0.979$ \\
 \enddata
\tablenotetext{a}{2MASS All-Sky Point Source Catalog \citep{https://doi.org/10.26131/irsa2}}
\tablenotetext{b}{Point sources from a Spitzer/IRAC survey of the Galactic center \citep{https://doi.org/10.26131/irsa505}}
\tablenotetext{c}{ALLWISE Source Catalog \citep{https://doi.org/10.26131/irsa1}}
\tablenotetext{d}{IRC Point Source Catalog: \url{https://www.ir.isas.jaxa.jp/AKARI/Archive/Catalogues/PSC/}}
\tablenotetext{e}{IRAS Point Source Catalog v2.1 (PSC) \citep{https://doi.org/10.26131/irsa4}}
\tablenotetext{f}{Hi-GAL Catalogs v1.0.2: \url{https://tools.ssdc.asi.it/HiGAL.jsp}}
\end{deluxetable*}

The center position of the slit nearly overlaps with 2MASS\,J17470424-2833204 (red $+$ in Figure~\ref{fig:image}) and is located by $\sim 1\farcs5$ south of the source 
SSTGC\,0753438 of the Spitzer IRAC Survey of the Galactic center 
\citep[blue $\times$,][] {2008ApJS..175..147R}.  
The IRAC color  $[3.6] - [8.0]$ of 4.5 suggests that the source is very red.
Wide-field Infrared Survey Explorer (WISE) and AKARI/IRC also detected a point source close to the position of the present object \citep[Table~\ref{tab:target},][]{2010A&A...514A...1I, 2010AJ....140.1868W}.  
There is also an IRAS source, IRAS\,17438-2832, which is separated by $26\farcs$9 in the west direction.  It is identified as AFGL\,2006 and indicated by the green $\times$ in Figure~\ref{fig:image}.  
The Herschel Infrared Galactic Plane Survey (Hi-GAL) also detected a source at 70\,$\mu$m, which was very close to the center position of the IRC slit \citep[white $+$,][]{2016A&A...591A.149M}.
The counterpart of AFGL\,2006 in the NIR (V8-5) was observed spectroscopically, which showed the presence of weak Br$\alpha$ and \ion{He}{1} emission \citep{1994MNRAS.268..194M}. 
The positions of these sources together with the fluxes in the catalogs are summarized in Table~\ref{tab:target}.  Except for IRAS\,17438-2832 or AFGL\,2006, the source positions agree within
$4\arcsec$.  Since there is no other very red, infrared-bright source in the vicinity, 
and since the listed fluxes are consistent with each other except for WISE W2 (4.6\,$\mu$m), these sources seem to be the same object.  
The flux of WISE band (W2) may include the radiation from the extended structure because of its large aperture.  The present observations took spectra of the southern part of the
extended emission around the object according to the IRAC image (Figure~\ref{fig:image}).

\section{Results and analysis} \label{sec:results}
\subsection{Observed spectra and continuum fitting} \label{res:cont}
\begin{figure*}[ht!]
\epsscale{0.86}
\plotone{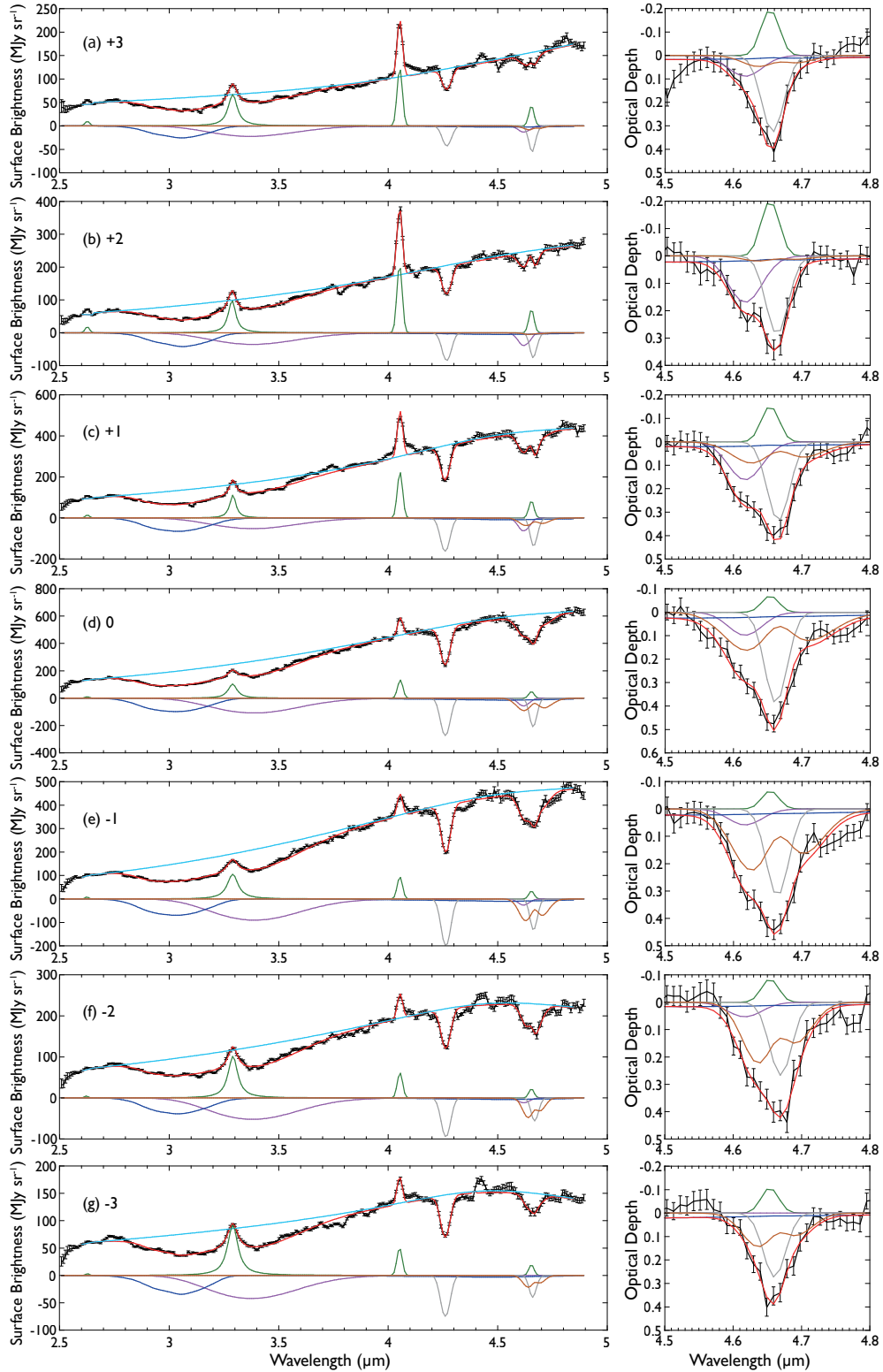}
\caption{The black lines show IRC spectra extracted at every 4$\farcs$38 in the slit.  Figures (a), (b), (c), ..., and (g) correspond to $+3, +2,$, ..., and $-3$ in the slit position shown 
in Figure~\ref{fig:image}, respectively.    In the left column, the red lines indicate the best fits, while the light blue lines show the assumed continua.  
The blue and orange lines show the absorption features of H$_2$O ice and CO gas, respectively.  The purple lines show the absorption of
the red wing at around 3.4\,$\mu$m and XCN at 4.62\,$\mu$m, while the gray lines indicate those of CO$_2$ ice at 4.26\,$\mu$m and
CO ice at 4.67\,$\mu$m.  The green lines show the emission features of PAH at 3.3\,$\mu$m and \ion{H}{1} recombination lines of Br$\alpha$, Br$\beta$, and 
predicted Pf$\beta$ (see the text).
The right column shows the optical depth of each component for 4.5--4.8\,$\mu$m.  The line colors are the same as in the left column.
See the text for details. \label{fig:spectra}}
\end{figure*}

The left column of Figure~\ref{fig:spectra} shows IRC spectra at seven positions from the west end ($+3$) to the east end ($-3$) of the slit.
Each spectrum clearly shows \ion{H}{1} Br$\alpha$ and Br$\beta$ emissions at 4.052 and and 2.626\,$\mu$m, respectively, and PAH emission at 3.3\,$\mu$m together with the broad absorption band
of H$_2$O ice at 3.0\,$\mu$m and the narrow absorption band of CO$_2$ ice at 4.26\,$\mu$m.   In addition, the
spectra indicate the presence of a broad, complex absorption feature at around 4.65\,$\mu$m, whose appearance varies with the position. 
The spectra also indicate excess emission at around 4.4\,$\mu$m over the continuum.

To derive the physical properties of the emission and absorption components, we fit the spectrum by several emission and absorption features.
First, we fit the continuum by a spline
function, as shown by the light blue lines in Figure~\ref{fig:spectra}.  The anchor points are chosen at 2.65, 3.96, 4.47, and 4.85\,$\mu$m, which avoid known gas and dust features.  
The fitted continuum implicitly includes the effect of the continuous foreground extinction,
$\tau_\mathrm{ext}(\lambda)$.  It is used to extract wide, strong ice absorption features and Br$\alpha$ emission in the observed spectra.  
To estimate the intensities of faint emission features, linear continua are assumed locally in a narrow spectral range because the spline continua may not give a sufficiently accurate baseline for the extraction of faint emission.
In the following, we describe the spectral fitting processes in each spectral region, separately.

\subsection{H$_2$O ice and PAH 3.3\,$\mu$m emission, and CO$_2$ Ice} \label{res:H2O}

The observed spectrum $F_\nu(\lambda)$ between 2.7 and 3.9\,$\mu$m is fitted with the following equation, assuming that the attenuating layers of the water ice and foreground
extinction are located in front of the region that emits the \ion{H}{1} lines and the 3.3\,$\mu$m feature:

\begin{equation}
F_\nu(\lambda) = (F_\mathrm{cont}(\lambda) + F_{3.3}(\lambda) ) \times \mathrm{exp}(-\tau_\mathrm{H_2O}(\lambda) - \tau_\mathrm{RW}(\lambda)), \label{eq:1}
\end{equation}

\noindent where $F_\mathrm{cont}(\lambda)$ is the fitted continuum, which is assumed to be $F_\mathrm{cont, int}(\lambda) \times \mathrm{exp}(-\tau_\mathrm{ext}(\lambda))$, with
$F_\mathrm{cont, int}(\lambda)$ being the intrinsic continuum corrected for the foreground extinction, and $F_{3.3}(\lambda)$ is the PAH 3.3\,$\mu$m emission, which
is approximated by a Lorentzian function. Note that $F_{3.3}(\lambda)$ thus defined also includes the effect of the foreground extinction $\tau_\mathrm{ext}(\lambda)$.
In Equation~(\ref{eq:1}), $\tau_\mathrm{H_2O}(\lambda)$ represents the water ice absorption, and $\tau_\mathrm{RW}(\lambda)$
takes account of its red-wing component.  For $\tau_\mathrm{H_2O}(\lambda)$, we adopt the optical constants of amorphous H$_2$O ice measured at various temperatures between 15 and 150\,K
\citep{2009ApJ...701.1347M}.  Because the 3.0\,$\mu$m band is broad and strong, we take account of the particle shape effect \citep{1997A&A...328..649E,
 2013ApJ...775...85N}.  We calculate the absorption cross section, assuming a continuous distribution of ellipsoids (CDE) with a peak at a sphere for the shape distribution \citep{1992A&A...261..567O}.
The red-wing component, $\tau_\mathrm{RW}(\lambda)$, accounts for the broad wing observed in various objects in the longer wavelength side of the H$_2$O ice absorption.
Its origin remains unclear at the present \citep[][references therein]{2013ApJ...775...85N, 2015ARA&A..53..541B}.   
In this study, the red wing is approximated by a Gaussian with a center wavelength of
3.35\,$\mu$m and a width (FWHM) of 0.42\,$\mu$m, which are determined from the fits at positions $-1$, 0, and $+1$.  
The best-fit temperature of H$_2$O ice ranges from 15 and 60\,K, suggesting no sign of crystallization of H$_2$O ice.  
The difference in the absorption profile of H$_2$O ice for temperatures 15--60\,K is small and does not affect the present results.  

The strong absorption feature at 4.26\,$\mu$m is attributed to CO$_2$ ice.  Its profile
is approximated by a single Gaussian because the intrinsic width is narrow \citep[18\,cm$^{-1}$;][]{2004ApJS..151...35G} compared to the present spectral resolution.
The low spectral resolution of the present spectra does not allow us to study the difference of the profile due to the environmental effects of CO$_2$ ice in detail \citep{1997A&A...328..649E,
 2013ApJ...775...85N}.
Br$\alpha$ emission at 4.052\,$\mu$m is fitted by a Gaussian.  
There is no sign of the emission features of aliphatic C\sbond H components at 3.4--3.5\,$\mu$m, which
usually appear together with the aromatic C\sbond H 3.3\,$\mu$m emission \citep{1989ApJ...341..278G}.
They might be masked by the broad red-wing component.   The observed ratio of the band intensities of the aliphatic C\sbond H to aromatic  C\sbond H components
ranges from 0.2 to 0.4 in \ion{H}{2}-PDR complexes \citep{2014ApJ...784...53M}, and the ratio of the peak intensities of the 3.4 to 3.3\,$\mu$m bands is expected to be in the range 0.1--0.3 with the
AKARI/IRC resolution.   CH$_3$OH ice features at $\sim 3.4$ and 3.54\,$\mu$m
\citep{1986A&AS...64..453D} are not recognized either.   An upper limit for the column density of CH$_3$OH ice is estimated as $1.3 \times 10^{17}$\,cm$^{-1}$, assuming that the band strength
of the 3.54\,$\mu$m band is $7.6 \times 10^{-18}$\,cm per molecule for pure CH$_3$OH ice \citep{1986A&AS...64..453D}.

The fit results are shown by the red lines in Figure~\ref{fig:spectra}.  Equation~(\ref{eq:1}) well fits the spectra of 2.5--4\,$\mu$m.  The CO$_2$ ice feature at 4.26\,$\mu$m is
fitted separately from Equation~(\ref{eq:1}).
Each component of the fits is shown by separate lines: the PAH 3.3\,$\mu$m and Br$\alpha$ emissions are shown by the green lines;
and the absorption of H$_2$O ice, its red wing, and CO$_2$ ice are indicated by the blue, purple, and gray lines, respectively. 

To estimate the column densities of the absorbing species, we adopt the values of the integrated band strength of H$_2$O ice obtained in \citet{2009ApJ...701.1347M}, while  
we simply assume the integrated band strength of the red-wing component as $20 \times 10^{-17}$ cm per molecule, which is equal to the typical band strength of the 3.0\,$\mu$m band of H$_2$O ice \citep{1995A&A...296..810G}, 
for the purpose of simple comparison.  Therefore, only the relative value is meaningful for the column density of the red-wing component given below.
The column density of CO$_2$ ice is estimated, assuming the band strength as $7.6 \times 10^{-17}$\, cm per molecule
\citep{1995A&A...296..810G}.  Since the absorption feature at 4.26\,$\mu$m is relatively deep and the intrinsic width is narrower than the present spectral resolution,
we take account of the saturation effect due to the low spectral resolution \citep{2010A&A...514A..12S}.  We simulate the feature convolving with the present resolution and estimate the
correction factor as described in \citet{2021ApJ...916...75O}.   The maximum correction factor applied is 1.1.
We note that the diffuse Galactic emission could affect the observed spectra of the present object because this is in the direction toward the Galactic center.  We discuss the effect in the next section
after estimating the foreground extinction.

\subsection{4.65\,$\mu$m feature and foreground extinction} \label{res:XCN}

The shape of the absorption feature at around 4.65\,$\mu$m varies from position to position and shows complex structures. We fit the spectra of 4.4--4.85\,$\mu$m with the following equation:

\begin{eqnarray} 
F_\nu(\lambda)  & = (F_\mathrm{cont}(\lambda) + F_\mathrm{Pf\beta} )  \times \mathrm{exp}(-\tau_\mathrm{CO ice}(\lambda) \nonumber \\
   &  - \tau_\mathrm{XCN}(\lambda)-\tau_\mathrm{CO gas}(\lambda)-\tau_\mathrm{H_2O}(\lambda)). \label{eq:2}
\end{eqnarray}

\noindent The strong absorption at the center is 
ascribed to the CO ice band at 4.67\,$\mu$m, which is represented by $\tau_\mathrm{CO ice}(\lambda)$. 
We approximate it by a single Gaussian because the intrinsic width is typically narrow \citep[$< 10$\,cm$^{-1}$;][]{2004ApJS..151...35G} compared to the present spectral resolution.
The environmental effects on the profile of CO ice are also difficult to be investigated with the present low-resolution spectra and because of other features superimposed.
The relatively broad width is attributed to 
CO gas absorption, $\tau_\mathrm{CO gas}(\lambda)$, which is often seen in YSOs \citep[e.g.,][]{2003A&A...408..981P, 2012A&A...538A..57A, 2021ApJ...916...75O, 2022ApJ...935..137K}.  
We calculate CO gas absorption with temperatures of 50, 100, 150, and 200\,K, convolving with 
the present spectral resolution, and search for the best-fit temperature.  Because of the low spectral resolution, we only make a rough estimate of the gas temperature from the broadness of the feature.
An asymmetric profile is clearly seen in the spectra of positions $+1$, 0, $-1$, and $-2$,
which suggests the presence of the XCN feature at 4.62\,$\mu$m.  
The XCN feature consists of two components, one with the peak at 4.60\,$\mu$m and the other with the peak at 4.62\,$\mu$m \citep{2005A&A...441..249V}.   They are not
clearly resolved with the present spectral resolution.  Therefore, we assume a single Gaussian with a center wavelength of 4.617\,$\mu$m and a width (FWHM) of 0.058\,$\mu$m for the XCN feature, $\tau_\mathrm{XCN}(\lambda)$.
The parameters are determined from the fits at positions $-1$, 0, and +1.
The contribution from the 4.5\,$\mu$m combination bands of H$_2$O ice, $\tau_\mathrm{H_2O}(\lambda)$, is included, which is extrapolated from the fits in the 3\,$\mu$m absorption feature.

The column density of CO ice is estimated, assuming the band strength as $1.1 \times 10^{-17}$\,cm per molecule \citep{1995A&A...296..810G}.  The saturation effect due to the low spectral resolution is also taken into account
using simulations in a way similar to CO$_2$ ice.  The correction does not exceed 7\%.  We assume the same band strength for both XCN components as $1.3 \times 10^{-16}$\,cm per molecule
to estimate the column density \citep{2005A&A...441..249V, 2011ApJ...740..109O}.  In the following, we refer to this as the column density of XCN for simplicity.  The XCN feature is weak and broad, and therefore, we do not apply correction for the saturation effect.

\begin{figure*}[ht!]
\epsscale{0.7}
\plotone{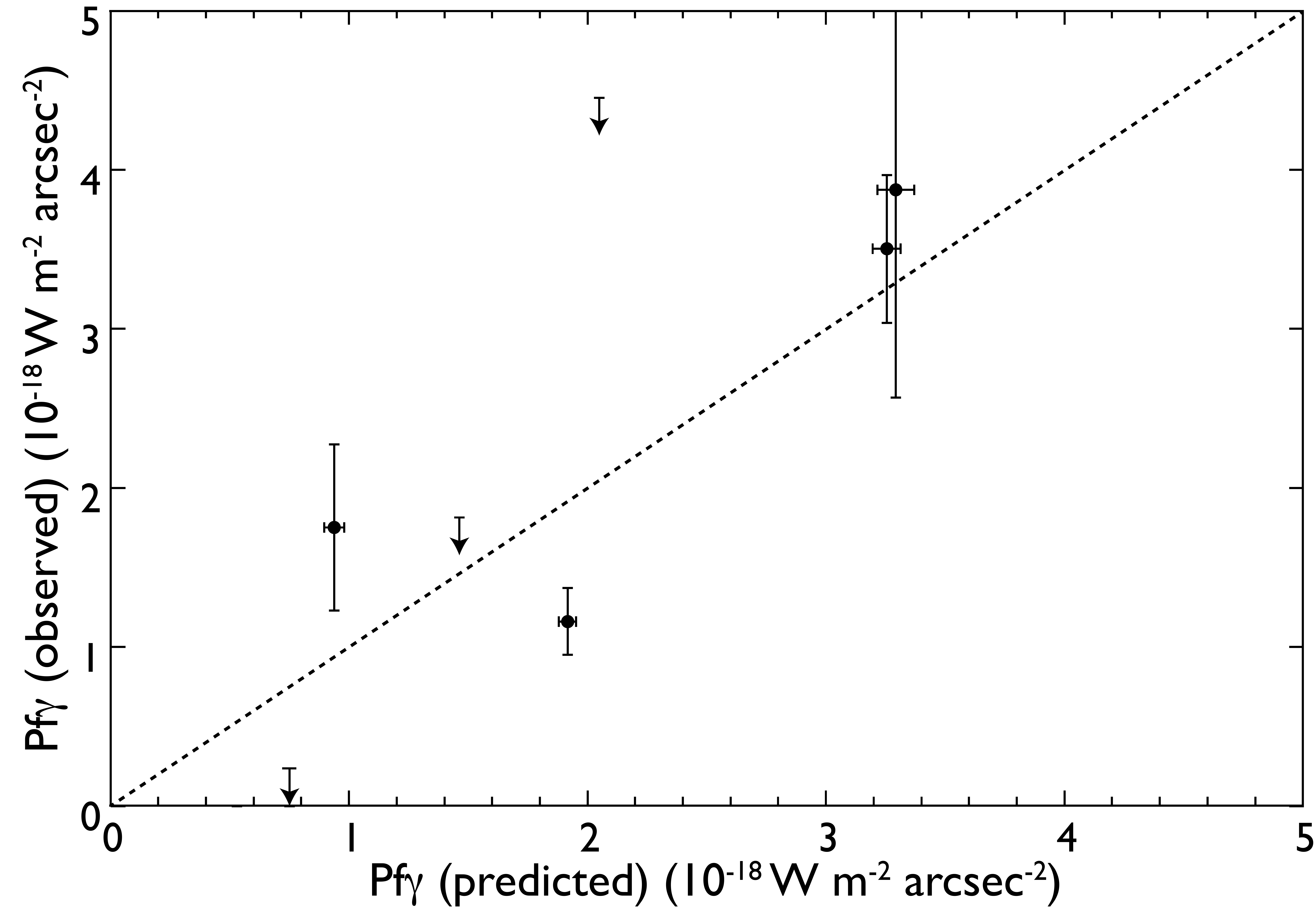}
\caption{Observed Pf$\gamma$ intensity at 3.74\,$\mu$m against the predicted intensity from the observed Br$\alpha$ intensity and the estimated extinction. 
The predicted intensity takes account of the foreground extinction as well as the extinction due to the red-wing component of H$_2$O ice.  The dashed line shows a line with a slope of unity.  
Only upper limits are given for the observed intensity at some positions. \label{fig:Pfg}}
\end{figure*}

In addition to these absorption features, we expect emission of \ion{H}{1} Pf$\beta$ recombination line at 4.654\,$\mu$m, denoted by $F_\mathrm{Pf\beta}$ in Equation~(\ref{eq:2}), 
because of the presence of strong Br$\alpha$ emission in all the spectra,
although it is not resolved in the present spectra.
To estimate its intensity,
we first derive the amount of the foreground extinction from the observed intensities of Br$\alpha$ and Br$\beta$ emissions, assuming the case B conditions with the electron density of $10^4$\,cm$^{-1}$ and the temperature
of $10^4$\,K \citep{1995MNRAS.272...41S}.  
Since H I Br$\beta$ at 2.626\,$\mu$m is located outside of the spline fit to the continuum, we estimate the continuum locally, assuming a linear baseline between 2.6 and 2.65\,$\mu$m, and
derive the intensity of Br$\beta$.
We adopt the latest extinction curve in the infrared based on observations with the Spitzer Infrared Spectrograph \citep{2021ApJ...916...33G} and estimate the foreground extinction.
The adopted extinction curve at 2.5--5\,$\mu$m is well approximated by a power-law dependence on the wavelength with an index of $-1.474$.  
It is shallower than the recent NIR extinction curve derived based on the ground-based observations by \citet{2022ApJ...930...15D}, who also show that the slope of the NIR extinction
curve has significant variations depending on the line of sight and that the average power-law index is estimated as $-1.7$ with the steepest slope being $-2.2$.  
It should also be added that the adopted extinction curve is derived for the diffuse ISM and may not be applicable to dense clouds.  
However, systematic studies of extinction from optical to infrared have only a limited number of
sight lines that probe dense regions.  Thus, we estimate the effect of uncertainties in the extinction curve simply by comparing the present results with those with the power-law indexes of $-1.7$ and $-2.2$.
The results with the power-law index of $-1.7$ show that the change of the power-law index makes no more than 3\% difference in the estimated Pf$\beta$ and even those with $-2.2$ 
show no more than 7\% difference because of the narrow spectral range in question.

Although it is faint, Pf$\gamma$ emission at 3.74\,$\mu$m is also detected at several observed spectra.  We compare the observed intensity of Pf$\gamma$ with that predicted from the observed intensity of 
Br$\alpha$ and the estimated extinction, to check the reliability of the estimated extinction.  The foreground extinction, $\tau_\mathrm{ext}(\lambda)$, as well as the extinction due to the red-wing component of 
H$_2$O ice, $\tau_\mathrm{RW}(\lambda)$, are taken into account in the prediction of the Pf$\gamma$ intensity.  The red wing makes only 10\% correction and does not make a significant effect.
Figure~\ref{fig:Pfg} shows the comparison, which suggests that they are in fair agreement and confirms that the estimated extinction is not largely in error.  
Note that $\tau(\mathrm{Br}\alpha) = 1$ corresponds to $A_V \sim 22$\,mag according to the adopted extinction curve.  
The derived $\tau(\mathrm{Br}\alpha)$ ranges 1.3--2.0 (see Figure~\ref{fig:position}(b)), suggesting $A_V \sim 29-44$\,mag in the observed region.  As noted above, the adopted extinction curve is derived for the diffuse ISM.
In denser regions, the optical to NIR extinction ratio could be smaller \citep{2021ApJ...916...33G, 2022ApJ...930...15D}.   Therefore, the estimated $A_V$ is only indicative and should be regarded as an upper limit.

\begin{figure*}[ht!]
\epsscale{0.5}
\plotone{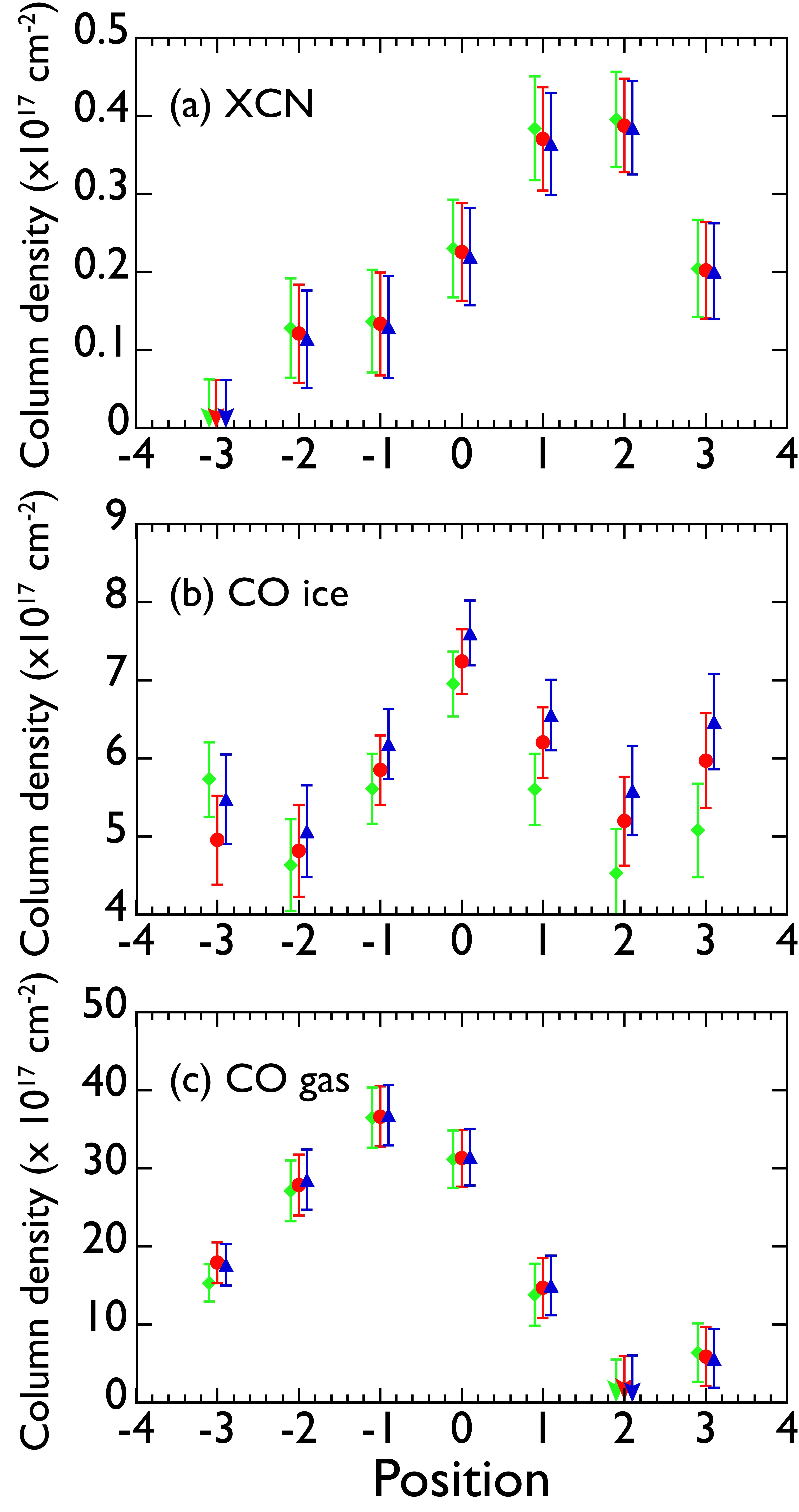}
\caption{Column densities of XCN, CO ice, and CO ice along the slit for three cases of the Pf$\beta$ intensity estimate. The estimates of the column densities of (a) XCN, (b) CO ice, and (c) CO gas are plotted against the slit position.
The standard cases are shown by the red circles, while the results of no extinction correction (minimum Pf$\beta$) are indicated by the green diamonds.   The maximum Pf$\beta$ cases
(addition of 3$\sigma$ uncertainty) are indicated by the blue triangles.  The results of the three cases are slightly shifted horizontally to avoid overlapping. \label{fig:comp}}
\end{figure*}

The fitted spectra are shown by the red lines in Figure~\ref{fig:spectra}.  They are also shown in the enlarged figures of the right column 
in optical depth units.  The assumed Pf$\beta$ emission is indicated by the green lines, and the absorption of CO ice, CO gas, and XCN is indicated by the gray, brown, and purple lines, 
respectively.  The blue lines indicate the absorption due to H$_2$O ice.
Equation~(\ref{eq:2}) fits all the observed spectra for 4.4--4.85\,$\mu$m fairly well.
Note that, at position $-3$, the XCN component is not detected with a significant level, and CO gas is not detected significantly at positions +2 and +3.
The combination of the absorption of CO ice, XCN, CO gas, and the Pf$\beta$ emission well accounts for the observed complex appearance of the broad feature at around 4.65\,$\mu$m, 
particularly the double-peak absorption feature at position +2.
The varying appearance of the feature along the slit can be attributed to the difference in the relative contributions from these components.

While the inclusion of Pf$\beta$ emission makes a fit significantly better, it is not resolved in the present spectra.  We carefully check the uncertainties associated with the assumed Pf$\beta$ intensity.
We fit the spectra without extinction correction, which gives the minimum intensity, as well as with the case where Br$\alpha$ intensity is increased by 3$\sigma$ of the measurement uncertainty,
to simulate a maximum intensity.  1$\sigma$ uncertainty for the Br$\alpha$ intensity is typically about 5\%.  Therefore, in the latter case, we increase the Br$\alpha$ intensity roughly by 15\%.
The results are shown in Figure~\ref{fig:comp}.
As shown in the right figures of Figure~\ref{fig:spectra}, the position of Pf$\beta$ is close to the absorption features of CO ice, and thus the uncertainty of Pf$\beta$ directly affects the column density of
CO ice.  However, the effect is still about 10\%, being comparable to the uncertainty in the estimate of the CO ice column density.  The effect on the XCN feature is rather indirect, and small compared to the uncertainty in the fits.
The effect on the CO gas component is also small, because the column density is determined mostly by the wing of the feature.

\ion{H}{1} Pf$\delta$ at 3.297\,$\mu$m is also expected to be superimposed on the PAH 3.3\,$\mu$m emission, but
not resolved in the present spectral resolution.  Using the derived extinction and the observed intensity of Br$\alpha$ line, 
we estimate its contribution to the 3.3\,$\mu$m emission as 4\% at maximum, which is comparable to the measurement uncertainties.  
It is also independently confirmed by the observed
intensity of Pf$\gamma$, which must be stronger than Pf$\delta$, that the contribution of Pf$\delta$ to the 3.3\,$\mu$m emission is small.  
In the following, the expected contribution from Pf$\delta$ is subtracted from the observed PAH 3.3\,$\mu$m intensity.  

As noted in section~\ref{res:H2O}, there should be PAH 3.3\,$\mu$m and continuum emission from the diffuse ISM, which could be significant \citep{1988A&A...201L...1G, 1996PASJ...48L..53T, 2013PASJ...65..120T}.
We extract a spectrum at the Ns slit of the width of $5\arcsec$ of the IRC \citep{2007PASJ...59S.401O} at a position $\sim 2\arcmin$ west of the present target in the same observation to estimate the contribution  
from the diffuse ISM.  The 3.3\,$\mu$m emission is barely seen in the spectrum, and its intensity is estimated as $\sim 8 \times 10^{-18}$\,W\,m$^{-2}$\,arcsec$^{-2}$.  This is about twice of the intensity obtained near the Galactic plane reported by
\citet{2013PASJ...65..120T}.  
Equation~(\ref{eq:1}) implicitly assumes that the diffuse radiation is located behind the absorbing layer.  In this case,
the contribution to the PAH 3.3\,$\mu$m emission from the diffuse ISM is only 4\% at most. 
Part of the emission may be present in the foreground.  Even if all of the 3.3\,$\mu$m emission in the diffuse ISM comes from the foreground of the target, it would be 35\% of the faintest intensity at position +3.  
The contribution of the foreground diffuse emission would make 
a constant offset in the derived 3.3\,$\mu$m emission intensity, but the general trend of the spatial variation should not be affected. 

\begin{figure*}[htb!]
\epsscale{1.15}
\plotone{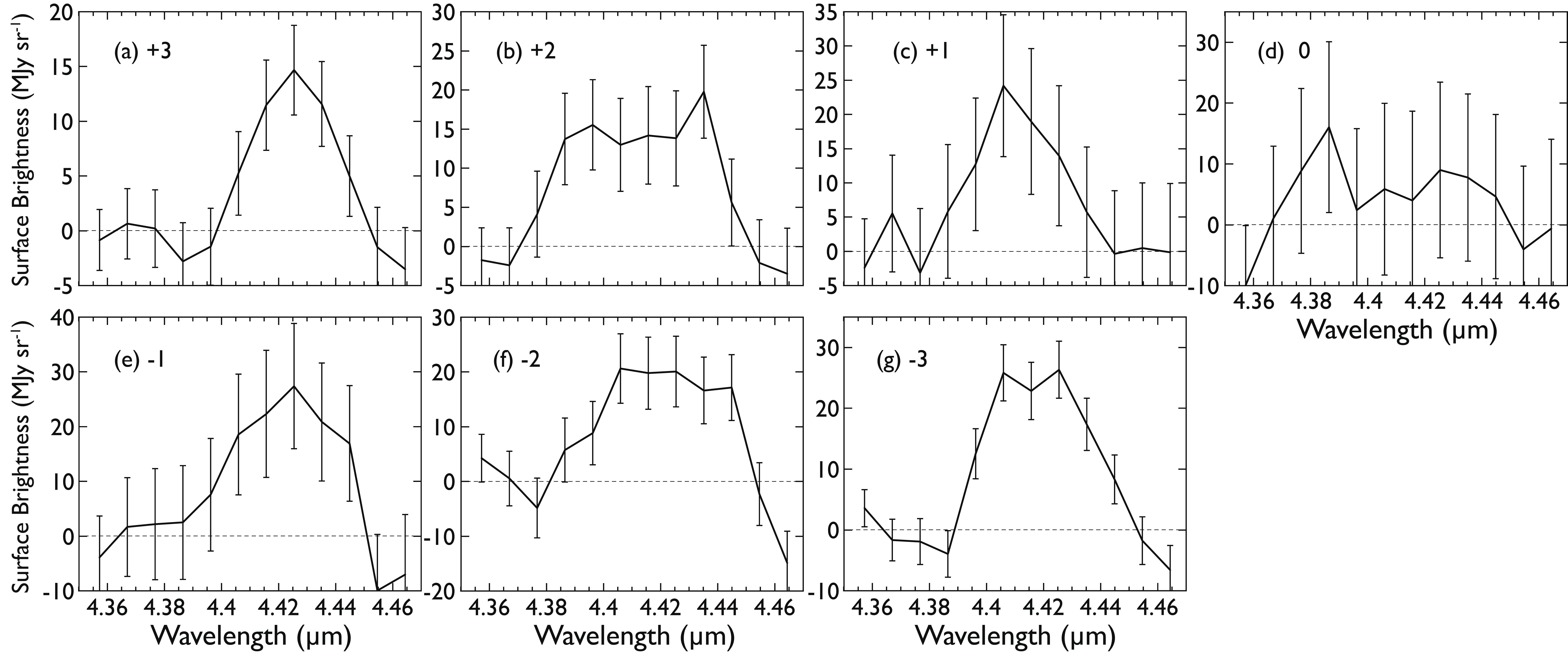}
\caption{Enlarged portion of the 4.4\,$\mu$m region spectra to show the 4.4\,$\mu$m excess after subtracting a linear baseline between 4.37 and 4.45\,$\mu$m. At position 0, the excess
is not detected with a significant level. \label{fig:C-D}}
\end{figure*}

The foreground diffuse continuum emission could also fill in the absorption features and weaken them.  The background level of the region around the present observations is estimated as 15 and 20 \,MJr\,sr$^{-1}$ at IRAC band 1 (3.6\,$\mu$m) and band 2 (4.5\,$\mu$m)
using the images of IRAC.  If we assume half of the emission coming from the foreground as a worst case, the diffuse emission would reduce the optical depths of H$_2$O ice by 20\%
and CO$_2$ ice by 10\% at position $-3$, where the observed continuum is the faintest among the observed region.  At other positions, the continuum is brighter, and the effect is smaller.  In this case, 
the abundance of CO$_2$ ice would increase by 10\% at most
due to the effect of the foreground diffuse continuum emission.  Throughout the present paper, the abundance of the species is given relative to H$_2$O ice.
The effect of the foreground diffuse emission on the column densities of XCN and CO ice is estimated as the same as for CO$_2$ ice since
the continuum intensity is similar.  It would reduce the column densities by 10\% and increase their abundance by 10\%.

\subsection{4.4\,$\mu$m excess and other features} \label{res:CD}

In addition to the above features, an excess emission at around 4.4\,$\mu$m, which can be attributed to the aromatic C\sbond D stretching vibration, is seen with a significant level at every observed position except for position 0 (Figure~\ref{fig:spectra}).  
There is also excess emission at around 4.8\,$\mu$m in the spectrum at position +3.  A similar excess emission may be hinted at positions +1 and $-1$, but the detection level is not significant.
It is located at the longest wavelength edge of the range expected for aliphatic C\sbond D features 
\citep{2020ApJ...892...11B, 2021ApJ...917L..35A} and is longer than those observed in other PDRs \citep{2004ApJ...604..252P, 2014ApJ...780..114O, 2016A&A...586A..65D}.
Its origin is not clear at present.  We will not discuss it in the present paper.

We estimate the intensity of the 4.4\,$\mu$m feature robustly by integrating the emission above a linear baseline between 4.37 and 4.45\,$\mu$m because it is
weak, and because its profile is not well defined.   The baseline points are chosen to extract the excess emission properly and avoid contributions from other features as much as possible.  
Figure~\ref{fig:C-D} shows the continuum-subtracted spectra of the region 4.36--4.46\,$\mu$m.
We note that there might be absorption of $^{13}$CO$_2$ ice at 4.38\,$\mu$m, which could affect the estimate of the baseline.  We estimate its contribution using the $^{12}$CO$_2$ ice absorption at 4.26\,$\mu$m.  
The observed ratio of the column densities of $^{13}$CO$_2$ to $^{12}$CO$_2$ ices ranges 0.009--0.02 \citep{2000A&A...353..349B}.
Here, we assume that the column density ratio is 0.02 as a conservative estimate and adopt $7.8 \times 10^{-17}$\,cm per molecule as the absorption strength of $^{13}$CO$_2$ ice \citep{1995A&A...296..810G}.
We estimate the absorption depth of $^{13}$CO$_2$ ice from the column density of $^{12}$CO$_2$ ice estimated, taking account the saturation effect. 
The results suggest that the contribution from $^{13}$CO$_2$  ice could overestimate the intensity of the 4.4\,$\mu$m excess by 10\% at most for most observed positions and that the largest value is found at position +3 as 20\%. 
These are still within the measurement uncertainties of the estimated intensity of the excess.  
The PAH emission is supposed to originate from the region adjacent to the ionized gas \citep{2008ARA&A..46..289T}, and the same foreground extinction is assumed to be applicable.  
In the following, the intensities of the C\sbond H emission at 3.3\,$\mu$m and C\sbond D emission at 4.4\,$\mu$m are corrected for the foreground extinction, $\tau_\mathrm{ext}(\lambda)$.  

The presence of the aromatic C\sbond D feature suggests the possible presence of the aliphatic C\sbond D features at 4.6--4.7\,$\mu$m, which may affect the analysis of the 4.65\,$\mu$m absorption feature.
The profile of the aliphatic C\sbond D feature is not well defined, but past detections suggest that it is broad \citep[$\sim$\,0.05\,$\mu$m,][]{2004ApJ...604..252P, 2014ApJ...780..114O, 2016A&A...586A..65D}.  
Thus, we assume that the effect on each component of the decomposition of the 4.65\,$\mu$m feature is not large.
It may lead to the underestimate of the CO gas absorption.  Further studies with higher spectral resolution are needed to clarify the contribution from the C\sbond D feature.

\subsection{Spatial variation of the column density and intensity} \label{res:spatial}

\begin{figure*}[ht!]
\epsscale{0.73}
\plotone{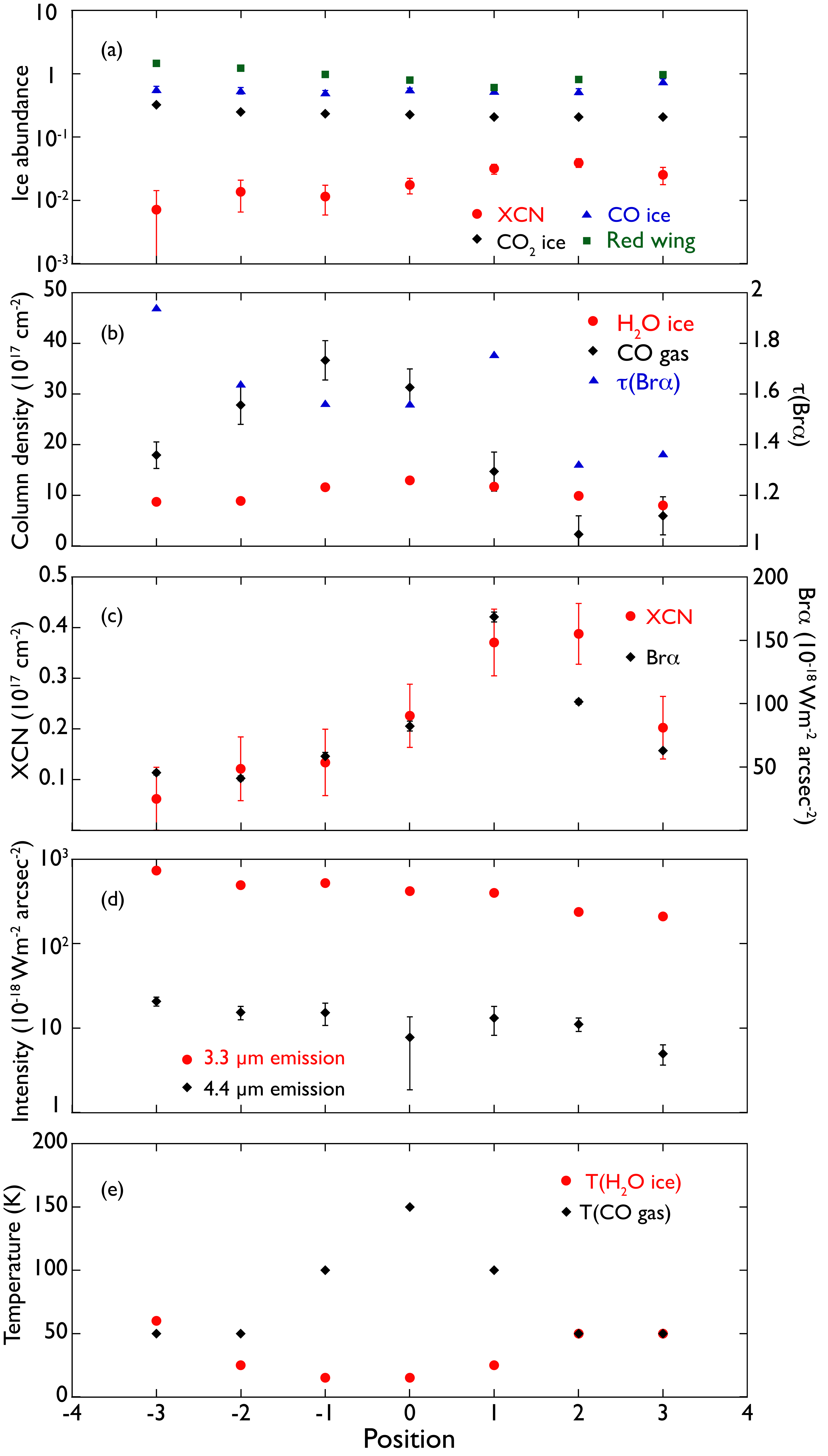}
\caption{Spatial variations of the column densities, abundances, intensities, and temperatures along the slit.  All the intensities of the emission features are corrected for the foreground extinction.  
(a) The abundances of the column densities of XCN, CO$_2$ ice, CO ice, and red-wing component  
are shown by the red circles, black diamonds, blue triangles, and green squares, respectively.  
(b) The column densities of H$_2$O ice and CO gas are indicated by the red circles and black diamonds, respectively (left axis),
while the extinction at Br$\alpha$ is shown by the blue triangles (right axis).
(c) The column density of XCN is shown by the red circles (left axis), and the intensity of Br$\alpha$ is indicated by the black diamonds (right axis).
(d) The intensities of the 3.3 and 4.4\,$\mu$m emissions are indicated by the red circles and black diamonds, respectively.
(e) The best-fit temperature of H$_2$O ice is shown by the red circles, and that of CO gas is indicated by the black diamonds.  \label{fig:position}}
\end{figure*}

\begin{figure*}[ht!]
\epsscale{0.7}
\plotone{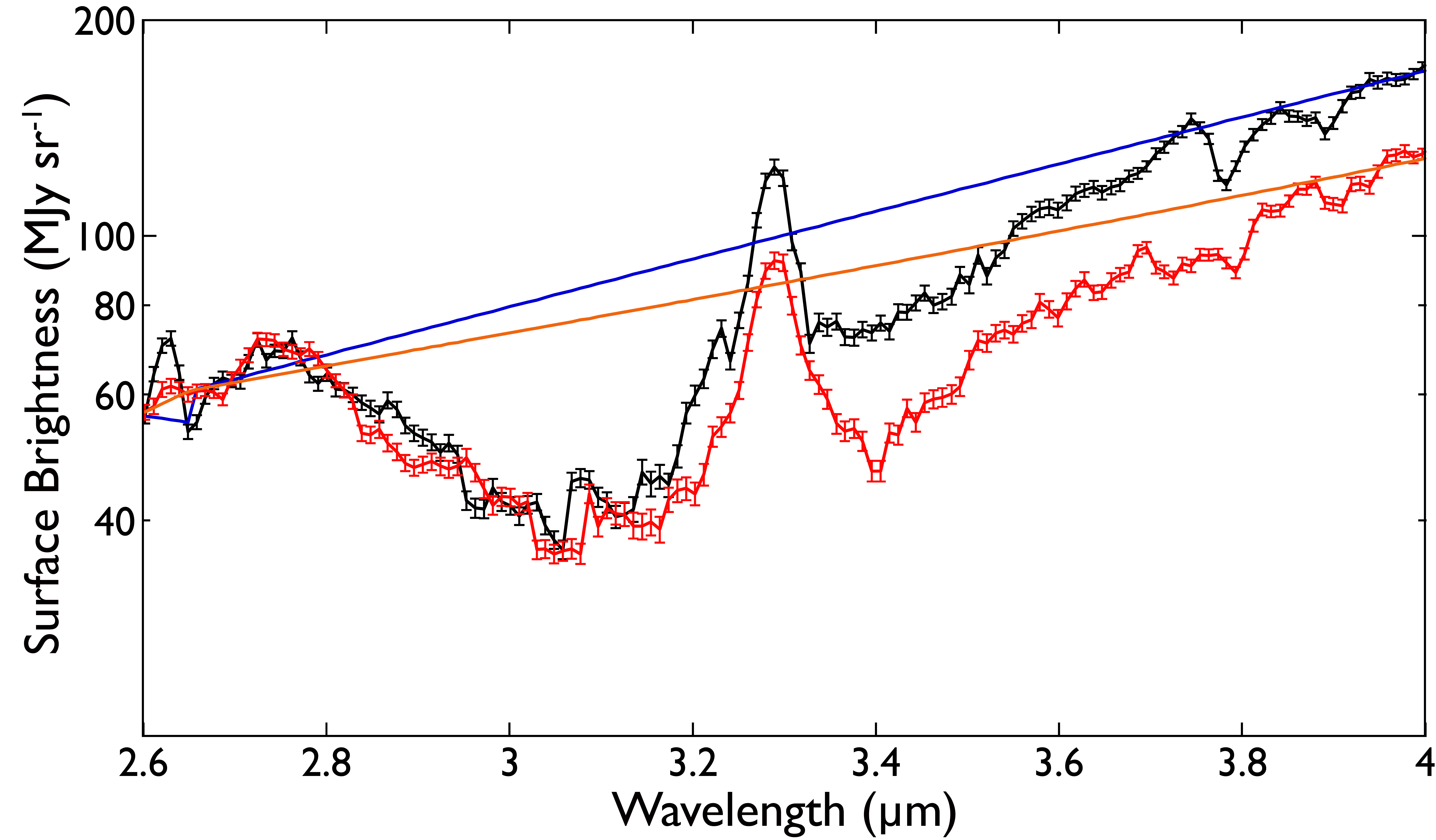}
\caption{The 3--4\,$\mu$m spectra of two extreme cases of the column density of XCN.  The black and blue lines show the observed spectrum and assumed continuum at
position +2, where the column density of XCN is the largest, while the red and orange lines indicate those at position $-3$, where XCN is not detected with a significant level. \label{fig:rw}}
\end{figure*}

Figure~\ref{fig:position} shows the spatial variation of the column densities, abundances, intensities of the emission bands, and temperatures of CO gas and H$_2$O ice along the slit.  
All the intensities of the emission features are corrected for the foreground extinction.
Figure~\ref{fig:position}a indicates the variations of the abundances of XCN, CO$_2$, CO ices, and red-wing component.  
The abundance of CO$_2$ ice (black diamonds) does not vary
largely along the slit, being 0.2--0.32.  The abundance of CO ice (blue triangles) stays also relatively constant as 0.50--0.57, except for position +3.  
The abundance of the red-wing component (green squares) shows a relatively large variation, from 0.6 to 1.5, suggesting that the red-wing is not correlated with H$_2$O ice.  
It shows a minimum abundance at position +1 and seems to show an anticorrelation with the XCN component.
To see the variation of the red-wing component with the XCN column density clearly, Figure~\ref{fig:rw} plots two extreme cases, position +2 (black line), where the XCN column density is the largest, and position $-3$ (red line), where the XCN feature is not
detected with a significant level, together with the assumed continua.  The column density of the red-wing component is primarily determined by the depth at around 3.4\,$\mu$m.  It is clearly seen that the depth of
the red wing is larger at position $-3$ than that at position +2.
Note that only the relative variation of the red wing is meaningful because we assume a fiducial band strength.

The XCN abundance shows a large variation, ranging 0.007--0.039 (red circles). The column density of the XCN component shows a good correlation with the intensity
of Br$\alpha$ (Figure~\ref{fig:position}c), which peaks at around positions +1 and +2.  This is notably different from the variations of the column densities of H$_2$O ice (Figure~\ref{fig:position}b) and other ice species.
The CO gas column density shows a peak at position $-1$, which is different from the column density of H$_2$O ice and the intensity of Br$\alpha$.
The excess at 4.4\,$\mu$m shows a similar variation to the PAH 3.3\,$\mu$m emission (Figure~\ref{fig:position}d).   
The spatial variations of these emissions are also relatively close to those of the extinction at Br$\alpha$ and the column density of CO gas (Figure~\ref{fig:position}b).  
The best-fit temperature of H$_2$O ice does not vary largely, reaching a minimum at the center of the slit and increasing toward the edges of the observed region.  Since the profile of H$_2$O ice does not
vary largely for 15--60\,K, the variation is not significant.   The temperature of the CO gas has a peak at position 0 
(Figure~\ref{fig:position}e), which is suggested by the broad width of the feature.  It suggests that the gas may be heated by the central source in its vicinity.

\section{Discussion} \label{sec:discussion}
\subsection{XCN feature and formation process of OCN$^-$} \label{dis:XCN}

\begin{figure*}[ht!]
\plotone{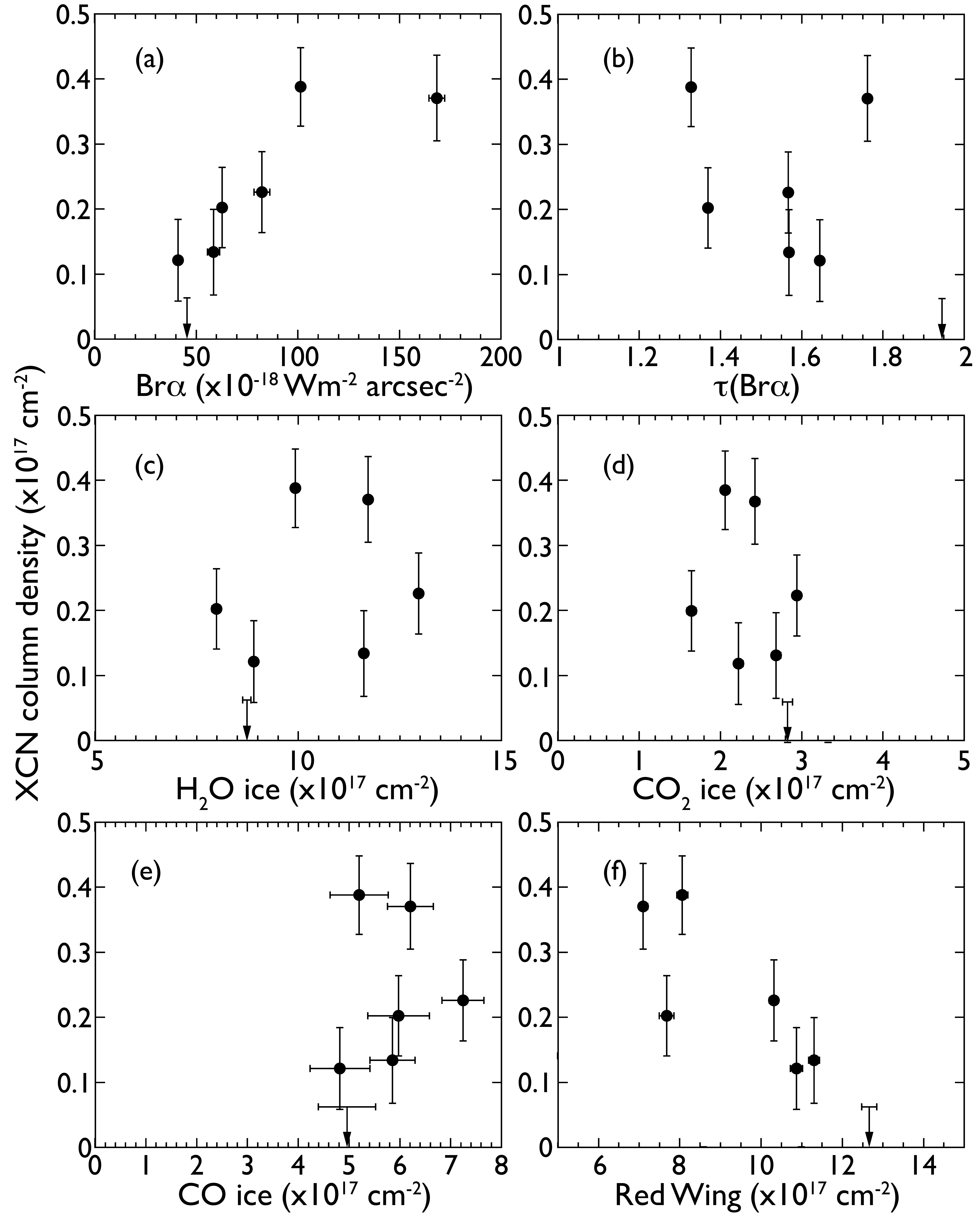}
\caption{Correlations of the XCN column density with (a) the Br$\alpha$ intensity corrected for the extinction, (b) the extinction at Br$\alpha$,
(c) the column density of H$_2$O ice, (d) the column density of CO$_2$ ice, (e) the column density of CO ice, and (f) the column density of the red-wing component.  
Note that the column density of the red-wing component is only relative since
the band strength is not known.  \label{fig:XCN-cor}}
\end{figure*}

Figure~\ref{fig:XCN-cor} shows correlations between the XCN column density and various parameters.  As expected from Figure~\ref{fig:position}, the XCN shows a good correlation with the Br$\alpha$ emission
(Figure~\ref{fig:XCN-cor}a).  The Pearson correlation coefficient is 0.85.  Such a correlation is indicated for the first time.  As discussed in section~\ref{res:XCN}, the estimate of the XCN column density is not
much affected by the uncertainty in the estimation of the Pf$\beta$ intensity.  Since CO ice, whose column density estimate should be more affected by Pf$\beta$, does not show a similar correlation,
the correlation between XCN and Br$\alpha$ should be secure and real.  
The correlation suggests that ionizing ultraviolet (UV) radiation enhances the formation of XCN \citep[e.g.,][]{1998A&A...339..553D, 1999ApJ...513..294P}.
The maximum abundance of XCN is found as $0.039 \pm 0.06$ at position +2, which is close to the maximum intensity of the Br$\alpha$ emission.
This abundance is higher than the largest value of 0.022 in W\,33\,A so far found in YSOs \citep{2000ApJ...536..347G}.  The large abundance is also found toward the infrared source in the Galactic center 
infrared source (IRS)\,19 as 0.032,
which has been corrected for the difference in the adopted band strength of XCN \citep{2002ApJ...570..198C} and in the central region of the nearby starburst-active galactic nulceus galaxy NGC\,4945 as 0.034--0.039 \citep{2003A&A...402..499S}.
The present maximum abundance is in a similar range to those in IRS\,19 and NGC\,4945, even if we take account of possible contamination by the foreground diffuse radiation.  
It should be noted that the present data do not
resolve the two components of the XCN feature.  Only the component peaking at 4.62\,$\mu$m is securely attributed to OCN$^-$ \citep{2004A&A...415..425V}.
While the carriers of the other component with a peak at 4.60\,$\mu$m are less clear, the correlations with other ice species suggest that both components are ascribed to OCN$^-$ \citep{2011ApJ...740..109O}.
The abundance of the 4.60\,$\mu$m component is generally small, being 0.012 at maximum \citep{2005A&A...441..249V}.  
The present XCN feature is best fitted with a Gaussian with a peak wavelength of 4.617\,$\mu$m.
While it needs to be confirmed by higher spectral-resolution
spectroscopy, the large abundance derived in the present results comes most likely from the 4.62\,$\mu$m component.  Note that the abundance of the 4.60\,$\mu$m component is less than 0.4\% of
OCN$^-$ in W\,33\,A \citep{2005A&A...441..249V}.  In the following discussion, we assume that the derived XCN column density represents that of OCN$^-$.

Originally, the XCN feature was thought to be a tracer of energetic processing of ices in YSOs \citep[e.g.,][]{1999ApJ...513..294P, 2001ApJ...550.1140H}.
Later, \citet{2004A&A...415..425V, 2005A&A...441..249V} show that UV photoprocessing of ice species has a maximum yield of the OCN$^-$ abundance of 0.027 based on the experiment of a mixture of
H$_2$O/CO/NH$_3$ = 100/53/32, which thus requires at least the NH$_3$ ice abundance of 0.3.
They argue that UV photoprocessing cannot account for the large abundance observed in W\,33\,A (0.022), because the estimated abundance of NH$_3$ ice is less than 0.05 \citep{2003A&A...399..169T}. 
They alternatively propose that thermal processing of acid-base chemistry is a more likely process to form OCN$^-$
\citep[see also][]{1998A&A...339..553D,  2003CP....288..197R, 2003CPL...368..594R}.  The yield of OCN$^-$ is limited by the abundance of the reactants. 
HNCO is proposed as a likely precursor to be converted into OCN$^-$ in the ice mantle.  However, it has not been detected in the solid phase; though it has been detected in the gas phase \citep{1972ApJ...177..619S}.
NH$_3$ is thought to be a possible electron donor, but its observed abundance also constrains the yield of OCN$^-$ to be formed \citep{2003A&A...399..169T,2004A&A...415..425V}.
Contributions from other reactants could change and increase the yield of OCN$^-$.
Later laboratory experiments show that there are several pathways to form ONC$^-$ in interstellar ices
\citep[e.g.,][]{2010ApJ...723..641B, 2016MNRAS.460.4297F}, and the actual formation process of OCN$^-$ is not fully understood yet.  

The understanding of the formation process of OCN$^-$ is important for the formation of prebiotic matter, such as amino acids,
in the ISM \citep[e.g.,][]{2015MNRAS.446..439F}.  Amino acids have recently been discovered in the sample of the asteroid Ryugu \citep{EizoNAKAMURA2022PJA9806B-01}, and their formation
process in the solar system is of great importance in the  astrobiological context.  
Amino acids in asteroids should have been formed in the cold, outer region of the solar system, which could have a direct connection to the formation of OCN$^-$ in the ISM.
The formation process of OCN$^-$ may also have implications on the formation of small nitrogen-bearing PAHs discovered in dense clouds
\citep{2021NatAs...5..181B, 2021Sci...371.1265M} as well as those suggested to explain the observed emission features in the mid-infrared by recent studies
\citep{2021ApJ...917..103E, 2021ApJ...923..202R, 2022PASJ...74..161V}.  

The present results show that the XCN abundance
has a good correlation with the intensity of \ion{H}{1} recombination line, providing the first clear evidence that UV photolysis plays an important role in the formation of the carriers of the XCN feature.  
They do not exclude the possibility of the XCN formation at the prestellar phase via energetic or nonenergetic processes.
UV photolysis was also thought to be effective in the formation of CO$_2$ ice \citep[e.g.,][]{2002ApJ...567..651W}.  The present data do not show any correlation of CO$_2$ ice with Br$\alpha$, suggesting that
XCN has a quite different formation path from other ice species, including CO$_2$ ice.
High UV flux also heats up the ice species and could enhance the thermal process.  However, the present spectra do not show
any sign of the crystallization or high-temperature characteristics ($> 60$\,K) of H$_2$O ices at around the peak position of the XCN column density.
The presence of CO gas suggests the heating of CO ice, but there is no apparent correlation of CO gas and ice abundances (Figure~\ref{fig:position}),
suggesting that the dominant portion of CO gas exists in a different region from the ice species.

The XCN column density does not show a clear correlation with the foreground extinction (Figure~\ref{fig:XCN-cor}b).  It does not correlate with H$_2$O, CO$_2$ and CO ices, either (Figure~\ref{fig:XCN-cor}c--e).  
Instead, the XCN column density shows a negative correlation
with the red-wing component (Figures~\ref{fig:rw} and \ref{fig:XCN-cor}f).  The Pearson correlation coefficient is $-0.79$.  The origin of the red wing is not well understood \citep{2015ARA&A..53..541B}.   
The present spectra show a very broad, smooth profile of the red-wing component and do not indicate any structures related to either hydrocarbons or CH$_3$OH ice seen toward the sources in the Galactic center
\citep{2002ApJ...570..198C, 2004A&A...425..529M, 2022ApJ...930...16J}. 
It may be an evolutionary indicator of both dust and ice mantle properties \citep{2013ApJ...775...85N}.  No clear correlation with H$_2$O ice and a very broad width found in the present study 
suggest that large-size H$_2$O ice cannot account for its characteristics, and the red wing is a separate component from H$_2$O ice.  
As the anticorrelation with XCN suggests, the red wing is also anticorrelated with Br$\alpha$.  
If the anticorrelation is directly linked to the formation of XCN, it may suggest that UV photolysis converts the carriers of the red wing
in the ice mantle into a mother species for the XCN formation.  
Further studies of the correlation in more objects are certainly needed to draw a clear conclusion. 

It should be added that the present spectra do not resolve each component of the features at 4.65\,$\mu$m (CO ice, CO gas, XCN, and Pf$\beta$), while Equation~(\ref{eq:2})
accounts for the general profile and its variation reasonably well, and the contribution from each component is reliably determined by the present analysis.  The spectra do not resolve the two components of XCN, either.
Further studies with
higher spectral resolution are definitely needed to obtain unambiguous conclusions on the XCN or OCN$^-$ formation processes.

\subsection{4.4\,$\mu$m excess and deuteration of PAHs} \label{dis:CD}
\begin{figure*}[ht!]
\plotone{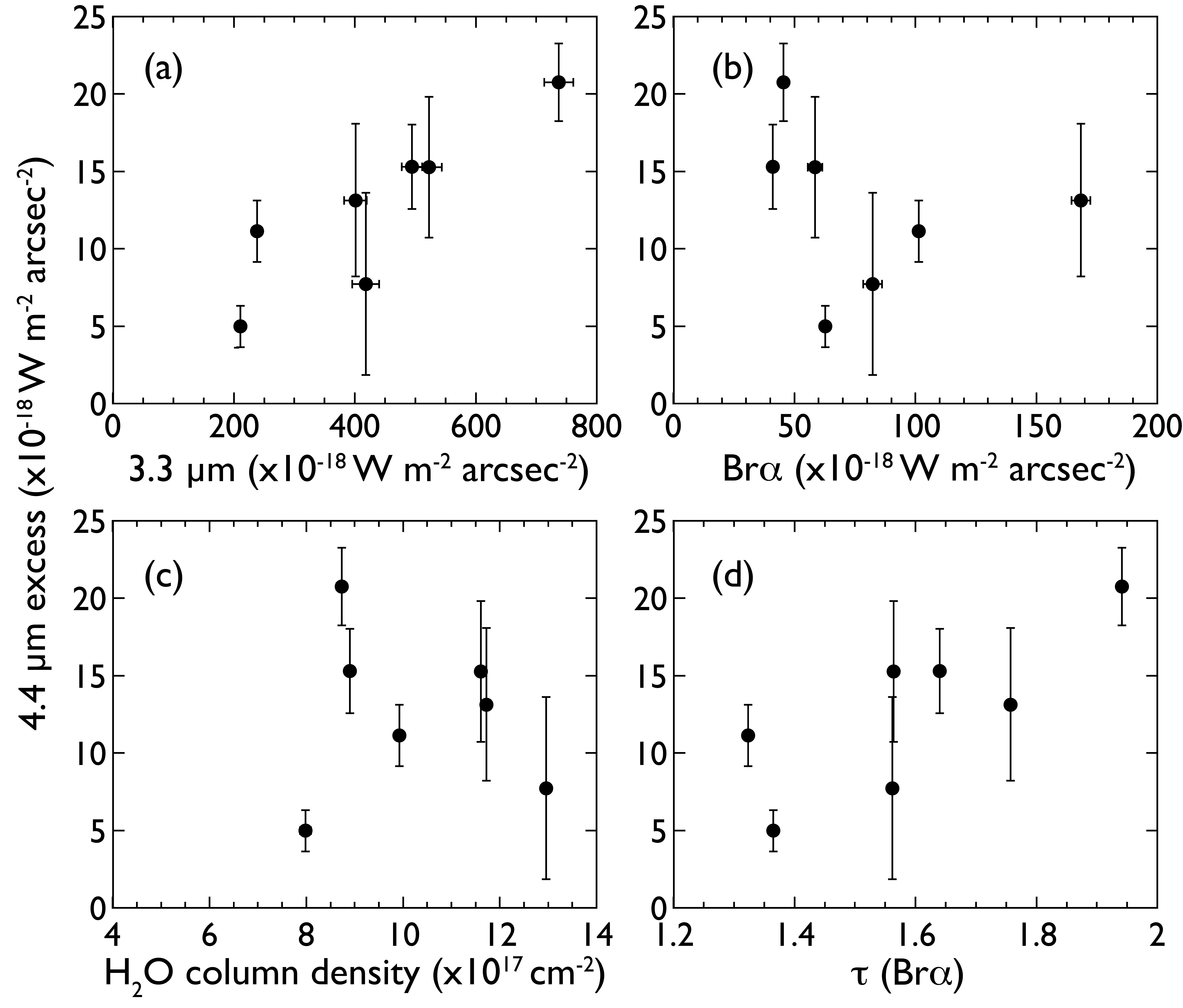}
\caption{Correlations of the 4.4\,$\mu$m excess emission with (a) the 3.3\,$\mu$m emission, (b) \ion{H}{1} Br$\alpha$,
(c) the H$_2$O ice column density, and (d) the extinction at Br$\alpha$.  The intensities of the 4.4 and 3.3\,$\mu$m emissions, and Br$\alpha$ are corrected for the foreground extinction. \label{fig:CD-cor}}
\end{figure*}

Figure~\ref{fig:CD-cor} plots correlations of the 4.4\,$\mu$m emission with several parameters.
The 4.4\,$\mu$m excess shows a positive correlation with the 3.3\,$\mu$m emission (Figure~\ref{fig:CD-cor}a).  The Pearson correlation coefficient is 0.87, suggesting that the correlation is good, although the number of the data is small.  
The correlations with Br$\alpha$ and H$_2$O ice are both weak with the correlation coefficient being $-0.22$ and $-0.12$, respectively (Figures~\ref{fig:CD-cor}b and c). The 4.4\,$\mu$m excess correlates also with the extinction at Br$\alpha$
to some extent (Figure~\ref{fig:CD-cor}d).  The correlation coefficient is 0.78.

Recent discoveries of small nitrogen-containing PAH molecules in dense clouds \citep{2021NatAs...5..181B, 2021Sci...371.1265M} suggest that there could also be nitrile C\tbond N stretch at around 4.46--4.50\,$\mu$m as the
4.62\,$\mu$m absorption feature is referred to as X\sbond C\tbond N \citep{2021ApJ...917L..35A}.  
The present excess is seen at wavelengths well shorter than the expected spectral range of C\tbond N stretch.  
\ion{H}{1} Br$\alpha$ emission peaks at position +1, while the excess emission peak is seen at position $-3$.  The excess emission does not show a clear correlation with Br$\alpha$ (Figure~\ref{fig:CD-cor}b).  
Therefore, it is also unlikely that the gas line emission originating from the ionized gas contributes to the excess emission.  The 3.3\,$\mu$m emission peaks at position $-3$ and varies nearly by a factor of 4 over the observed region.  
The correlation between the 4.4\,$\mu$m excess with the 3.3\,$\mu$m emission (Figure~\ref{fig:CD-cor}a) suggests that the 4.4\,$\mu$m excess seen in the present spectra is most likely to originate from the aromatic C\sbond D stretching vibration.

D is an element that was created in the first minutes after the Big Bang and has been monotonically decreased to the present by the nucleosynthesis in stellar interiors or astration.  
The abundance of D is thus important for the study of the nucleosynthesis in the early universe and the chemical evolution of our Galaxy \citep{1985ARA&A..23..319B, 2006MNRAS.369..295R}.  
Observations of distant quasars indicate that the primordial abundance of D relative to H is about 25 ppm \citep{2018ApJ...855..102C, 2018MNRAS.477.5536Z}.  
The abundance of D in the ISM, on the other hand, shows a large scatter without a clear correlation with the metallicity \citep{2006ApJ...647.1106L}, 
which cannot be explained solely by the chemical evolution of the Galaxy \citep{2010IAUS..268..153T}. 
A Bayesian analysis of the observed distribution of D suggests that the D/H ratio in the ISM ranges from $7\pm  2$ to $20\pm 1$\,ppm \citep{2010MNRAS.406.1108P}.  
The difference between the primordial value and the maximum D/H in the ISM
($\sim 5$\,ppm) can be attributed to the astration in the course of the evolution of the Galaxy.  The remaining fraction of 13\,ppm is missing in the ISM and thought to be depleted onto dust grains.  
A likely candidate that harbors the missing D in the ISM is suggested to be PAHs \citep{2006ASPC..348...58D}.  
Their aromatic units contain roughly 60\,ppm of the total H in the ISM at maximum \citep{2014ApJ...780..114O, 2022ApJ...933...35M} and could sequester the missing D
because of the difference in the zero-point energy.  Taking account of the abundance of PAHs, 
the D/H ratio of PAHs could be as large as 0.2--0.3 if PAHs sequester all the missing D in the ISM.

Excess emission at 4.6--4.8\,$\mu$m, which is attributable to the aliphatic C\sbond D stretch \citep{2020ApJ...892...11B}, has been detected with significant levels relative to the corresponding aliphatic C\sbond H features at 3.4--3.5\,$\mu$m 
in several PDRs.  However, the aromatic C\sbond D stretching feature at around 4.4\,$\mu$m has so far been detected with a level of 2--3\% at most, relative to the aromatic C\sbond H stretch at 3.3\,$\mu$m in its intensity, only in 
{a few \ion{H}{2} region--PDR complexes and a reflection nebula \citep{2004ApJ...604..252P, 2014ApJ...780..114O, 2016A&A...586A..65D}.  
Since aliphatic C\sbond H is not dominant relative to aromatic C\sbond H in the carriers that emit the emission features \citep{2016ApJ...825...22Y}, 
past observations cannot confirm if PAHs sequester a sufficient amount of D to account for the missing D in the ISM \citep{2014ApJ...780..114O, 2016A&A...586A..65D, 2021ApJS..255...23Y}.
Furthermore, past detections of the 4.4\,$\mu$m emission were made at a single position and did not allow us to study the spatial correlation with the C\sbond H stretch at 3.3\,$\mu$m within the object 
because either the observations were made with a large aperture \citep{2004ApJ...604..252P}, the spectra were uniform over the slit area \citep{2014ApJ...780..114O}, or the objects were bright only at a limited area in the slit \citep{2016A&A...586A..65D}.  
Since the feature is faint and the number of detected sources is small, it may not be straightforward to unambiguously assign the excess emission at 4.4\,$\mu$m to the aromatic C\sbond D stretching mode.  
The present study finds a correlation with the 3.3\,$\mu$m aromatic C\sbond H emission, providing clear evidence for the assignment of the observed 4.4\,$\mu$m emission to C\sbond D stretch for the first time.

The present data show that the ratio of the 4.4 to 3.3\,$\mu$m emission intensities peaks at position +2, being $0.047 \pm 0.009$.  
The average ratio in the observed region is $0.029 \pm 0.002$.  The intensities of both emissions are corrected for the foreground extinction.
If the extinction is not corrected, the peak and average ratios become $0.087 \pm 0.016$ and $0.059 \pm 0.006$, respectively.  
The largest ratio of intensities of the 4.4 to the 3.3\,$\mu$m emission so far obtained is $0.031 \pm 0.007$ at a position in IRAS\,12073-62337 \citep{2016A&A...586A..65D}.  
The present peak value is larger than this ratio, where the 4.4\,$\mu$m excess is detected with 5.6$\sigma$, and even the average value is close to it.  

The 4.4 to 3.3\,$\mu$m emission intensity ratio depends on the relative band strength and the difference in the excitation conditions.
The latest experiment confirms that the relative band strength of the stretching vibrations between aromatic C\sbond H and C\sbond D agrees with the predicted value 
and that the estimates of D/H in previous studies from the observed 3--5\,$\mu$m emissions are not affected \citep{2022ApJ...933...35M}.
Translation of the intensity ratio into the D/H ratio also depends on the parameters in the emission model because the 4.4\,$\mu$m band is expected to be excited more easily than the 3.3\,$\mu$m band by lower-energy photons 
\citep{2014ApJ...780..114O, 2016A&A...586A..65D, 2021ApJS..255...23Y}.  Therefore, the ratio of the excitation between 3.3 and 4.4\,$\mu$m bands depends on the spectrum of the exciting radiation and the size of PAHs.  
The present target is embedded in a dense envelope of a YSO, and the incident radiation may have less UV photons.  However, the presence of \ion{H}{1} recombination lines indicates that there should be a sufficient amount of UV photons 
to ionize the gas and excite the 3.3\,$\mu$m band.  Therefore, the excitation conditions may not differ significantly from other bright PDRs.  It is not straightforward to attribute the larger ratio to the difference in the incident radiation spectrum.

PAHs can be deuterated in several ways \citep{1997AIPC..402..523T, 2001M&PS...36.1117S}.  The gas-phase unimolecular photodissociation through scrambling enriches D in PAHs under UV radiation efficiently \citep{2020A&A...635A...9W}.  
This process does not require low temperatures and should be efficient only for small PAHs, which are responsible for the emission features at 3--5\,$\mu$m.  It occurs efficiently in dense regions and produces a large D/H ratio in PAHs ($\sim 0.15$).
In dense regions, cosmic rays produce UV radiation \citep{1997AIPC..402..523T} and the atomic D/H ratio can be very large ($\sim 0.05$) because H$_2$ is efficiently formed on the grain surface, while HD is not \citep{1983A&A...119..177T}.  
The scrambling also leads to the preferential loss of aliphatic C\sbond H over aromatic C\sbond D and increases aromatic C\sbond D \citep{2020A&A...635A...9W}, which is consistent with the absence of aliphatic C\sbond H features in the present spectra.

D fractionation can also proceed in the gas phase by ion--molecular reactions at low temperatures due to the lower zero-point energy of D than that of H \citep{1997AIPC..402..523T, 2001M&PS...36.1117S}.  
For the regions with the number density of $3 \times 10^4$\,cm$^{-3}$, it is expected to take PAHs $\sim 3 \times 10^5$\,yr to become deuterated and sequester 1\% of the elemental D \citep{1997AIPC..402..523T}.  
The time scale seems to be comparable to the age of the present object because it is already ionizing the surrounding gas. 

Since H$_2$O, CO$_2$, and CO ices are present in the envelope surrounding the object, the processes on the ice surfaces are also expected to proceed efficiently.  
Gas--grain reactions are known to be efficient to enrich D in molecules that can be hydrogenated \citep{1997AIPC..402..523T, 1997ApJ...482L.203C}.  
There is, however, no laboratory evidence that PAHs are hydrogenated by simple gas--grain interactions and this process may not be efficient to produce deuterated PAHs in dense regions \citep{2001M&PS...36.1117S}.  

\begin{figure*}[ht!]
\epsscale{0.8}
\plotone{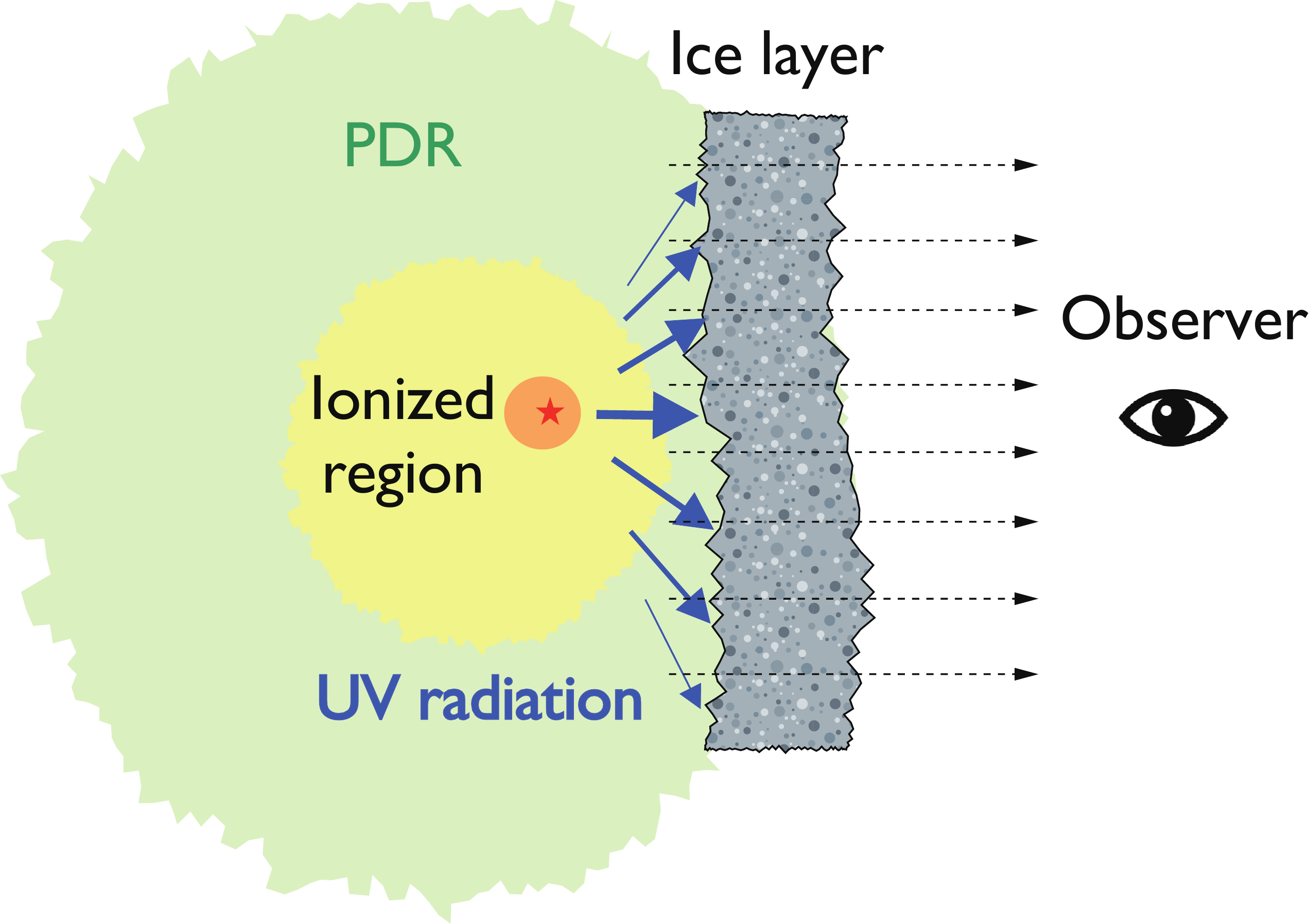}
\caption{Schematic view of the proposed geometry.  The ice layer has a slab-like shape (gray slab), separated from the central ionized region.
The ionized region has a compact ionizing source (red star in the brown region), which provides strong UV radiation, surrounded by an extended region \citep[yellow region,][]{2001ApJ...549..979K}.
The intensity of the UV radiation on the surface of the ice layer (blue arrows) varies with the distance from the central source, affecting the efficiency of the
formation of the XCN carriers.  The width of the arrow schematically indicates the intensity of the UV radiation at the surface of ice layer.  
The light green region is a PDR, where PAH and deuterated PAH emission arise.  The dashed arrow shows the light rays to the observer.
 \label{fig:model}}
\end{figure*}

UV photolysis reactions in ice mantles can form complex molecules from simple ones and propagate D-enrichment of starting molecules to more complex organic species.  
This process could be dominant in D-enrichment reactions of PAHs in ice mantles \citep{1999Sci...283.1135B, 2000ApJ...538..691S}.  
It requires the presence of ice mantles and UV radiation, both of which exist in the present object.  Therefore, D-enrichment of PAHs by UV photolysis reactions in ice mantles is likely to occur in the present object.  
It is expected to produce both aromatic and aliphatic C\sbond D bonds.  The present data suggest a correlation of the 4.4\,$\mu$m excess with the extinction (Figure~\ref{fig:CD-cor}d), 
supporting that the cold environments are crucial for the formation of aromatic C\sbond D.  However, any clear correlation of the intensity of the 4.4 or the 4.4 to 3.3\,$\mu$m intensity ratio is not seen with the column density of H$_2$O ice 
(Figure~\ref{fig:position} and \ref{fig:CD-cor}c).  Deuteration proceeds on ice surfaces in cold environments, while the release of PAHs and deuterated PAHs occurs at the surface of the ice layer from the grain surface.
Therefore, it may not be surprising that no direct correlation is seen with the H$_2$O ice column density.

The formation of aromatic C\sbond D bonds can be controlled by gas-phase reactions (ion--molecular reactions and unimolecular photodissociation), while aliphatic C\sbond D bonds may form most effectively via reactions on the grain surface \citep{2004ApJ...604..252P}.  
The present observations do not allow us to study the aliphatic C\sbond D features because of the overlapping broad absorption feature at 4.65\,$\mu$m.  
On the other hand, the present detection of the aromatic C\sbond D feature in the dense envelope around a YSO supports that D-enrichment proceeds efficiently in low-temperature, high-density regions.  
PAHs formed in these environments can be highly deuterated \citep{2021ApJ...917L..35A}.  
No significant detection of the aromatic C\sbond D feature in bright PDRs in past studies may result from the dilution by less deuterated PAHs produced in stellar outflows 
or chemical reactions at high temperatures that make PAHs equilibrate with the low atomic D/H ratio in the warm gas phase \citep{1997AIPC..402..523T}.  
The excitation of short-wavelength bands does not occur without strong UV radiation.  
The present object provides a unique place to produce C\sbond D bonds in PAHs efficiently and excite 3--5\,$\mu$m emission sufficiently by UV photons, enabling the detection of both features.  
More sensitive observations of this spectral range are needed for the study of deuterated PAHs in dense, cold regions.

\subsection{Nature of the object and other ice features}

The presence of \ion{H}{1} recombination lines and PAH 3.3\,$\mu$m emission together with the large XCN abundance indicates typical characteristics of massive YSOs (MYSOs)
embedded in a thick envelope.  \citet{2001ApJ...549..979K} show that a typical size of ultracompact \ion{H}{2} regions with extended emission ranges 0.05--1\,pc.  To make a rough estimate,
we simply take a median value or 0.2\,pc for the radius of the present ionized region, including the extended component}, and that it corresponds to the observed radius $\sim 15\arcsec$, 
then the distance is estimated as $\sim 3$\,kpc.  
By integrating the observed flux from 3.6 to 100\,$\mu$m (Table~\ref{tab:target}), we obtain a lower limit of the total luminosity as $\sim$$3 \times 10^3 L_\odot$, which is
roughly consistent with the nature of the MYSO.  Note that this is a very rough estimate and that the uncertainty is large.  It only implies that if the distance is a few kiloparsecs,
the observed properties can be consistent with the MYSO identification.

The present data show that CO$_2$ ice is well correlated with H$_2$O ice and  
that the CO$_2$ ice abundance does not change largely, being 0.2--0.32 in the observed region.
The average ratio of $0.24 \pm 0.04$ is in the boundary between the abundances of MYSOs and low-mass YSOs \citep[LYSOs;][]{2011ApJ...740..109O, 2015ARA&A..53..541B}.
The abundance of CO ice does not show a large variation, either, ranging 0.50--0.75.  It becomes largest (0.75) at the west end (position +3), 
where CO gas is not detected significantly.
The abundance of CO ice varies among YSOs, and   
the present large value is more consistent with LYSOs rather than MYSOs \citep{2011ApJ...740..109O, 2015ARA&A..53..541B}.

The relatively constant column densities of the ice species except for XCN and the nonsystematic variation of $\tau(\mathrm{Br}\alpha)$ along the slit may be explained if
the ices reside in a slab-like layer rather than a spherical shell surrounding the ionized region as schematically indicated in Figure~\ref{fig:model}.  
The UV source is located at an off-center position, closer to the dense region as suggested by \citet{2001ApJ...549..979K} for an ultra-compact \ion{H}{2} region.
This geometry explains the variation in the intensity of Br$\alpha$, which correlates with the incident UV radiation onto the surface of the slab of ice species.
The efficient formation of the XCN carriers, which show a distinct variation from other ices, occurs via UV photolysis near the surface of the ice layer close to the central source.  
The formation of other ice species proceeds rather homogeneously in the slab, being controlled by the local conditions.
The present observations probe not only the direction toward the central source but also various positions in its thick envelope.
The large abundance of CO ice at the western edge of the observed region (position +3) may be due to the very cold conditions at the region far from the central source.
The observed variations in the abundances of various ices suggest that the ice formation and/or processing are controlled by
 local physical conditions within the envelope. 

The observed ice abundances may also be affected by the thermal history of the object.
The central source should be in a stage of forming a compact \ion{H}{2} region, where CO ice is distilled from CO$_2$ ice \citep{2011ApJ...740..109O, 2015ARA&A..53..541B}.
The low spectral resolution of the present data, however, does not allow us to investigate the differences in the absorption profiles between the polar and apolar ices in detail, and we cannot
make a clear conclusion on the stage of ice formation.

\section{Summary} \label{sec:summary}

We present the results of NIR (2.5--5\,$\mu$m) long-slit ($\sim 30\arcsec$) spectroscopy of an envelope of a YSO in the direction toward the Galactic center with the IRC on board AKARI.
The present target is suggested to be the object previously designated as AFGL\,2006 based on its infrared characteristics and location, although the position does not match accurately.
We extract spectra at seven positions in the envelope.  The spectra show strong absorption features of H$_2$O and CO$_2$ ices at 3.0 and 4.26\,$\mu$m, respectively, and \ion{H}{1} Br$\alpha$ at 4.052\,$\mu$m, and
PAH 3.3\,$\mu$m emissions.
They also show a broad, complex absorption feature at around 4.65\,$\mu$m.  
We reproduce the feature successfully by a combination of absorption of CO ice, CO gas, and XCN, and Pf$\beta$ emission,
although the spectra do not resolve each component.
The present spectra do not resolve the two components of the XCN feature either, one of which is securely attributed to OCN$^-$.  
The XCN abundance becomes $0.039 \pm 0.06$ at maximum, which is comparable to the largest value so far found in NGC\,4945.
We found that the XCN component shows a distinct spatial variation from other ice species and well correlates with Br$\alpha$.  This suggests that UV photolysis plays an important
role in the formation of the carriers of the XCN feature for the first time.  
We also found that the broad red-wing component at around 3.35\,$\mu$m does not correlate with H$_2$O ice, but shows an anticorrelation with XCN.
It suggests that the carriers of the red wing are not related to H$_2$O ice and that they may play a role in the formation of XCN.
The present results have important implications on the formation processes of prebiotic matter in the ISM.

The present spectra also show excess emission at 4.4\,$\mu$m.  The excess shows a good correlation with the emission at 3.3\,$\mu$m, which is attributed to the stretching vibration of aromatic C\sbond H.
The correlation provides the first clear evidence that the 4.4\,$\mu$m emission comes from the aromatic C\sbond D vibration, which has been elusive in previous studies.  
The present data show the largest ratio of the 4.4 to 3.3\,$\mu$m emissions so far obtained.
The present detection suggests that the dense envelope with UV radiation would be crucial to detect the feature with a significant level and
that the deuteration of PAHs proceeds efficiently in cold, dense regions.

The presence of \ion{H}{1} recombination lines and 3.3\,$\mu$m PAH emission suggests that the object is likely an MYSO embedded in a thick envelope.
The ice species may reside in a slab-like layer rather than a spherical shell enclosing the central source.
The spatial variations of the ice abundances in the present object suggest that the formation and processing of an ice species may involve various processes sensitive to local physical conditions.
The present study is based on low-resolution spectroscopy, which does not allow us to resolve
individual components in the broad feature at 4.65\,$\mu$m and to study environments on which CO and CO$_2$ ices reside.  Observations with higher spectral resolution of the object
would certainly help elucidate the physical environments of ice species and their formation process, as well as the nature of the object.  Further searches of the C\sbond D feature in dense
regions will also be important for the understanding of the deuteration process of PAHs and for the hiding site of the missing D in the ISM.

\acknowledgments
This work is based on observations with AKARI, a JAXA project
with the participation of ESA.  The authors thank all the members of the AKARI project for their continuous support.  
This research has made use of the NASA/IPAC Infrared Science Archive, which is funded by the National Aeronautics and Space Administration and operated by the California Institute of Technology. 
This work was supported by JSPS KAKENHI grant Nos. JP18K03691, JP20H05845,  JP21H01145, and JP22H01261, as well as JSPS Bilateral Program grant No. 120219939.
%

\vspace{5mm} 
\facility{AKARI (IRC)}






\bibliography{XCN4}{}

\begin{thebibliography}{}
\expandafter\ifx\csname natexlab\endcsname\relax\def\natexlab#1{#1}\fi
\providecommand{\url}[1]{\href{#1}{#1}}
\providecommand{\dodoi}[1]{doi:~\href{http://doi.org/#1}{\nolinkurl{#1}}}
\providecommand{\doeprint}[1]{\href{http://ascl.net/#1}{\nolinkurl{http://ascl.net/#1}}}
\providecommand{\doarXiv}[1]{\href{https://arxiv.org/abs/#1}{\nolinkurl{https://arxiv.org/abs/#1}}}

\bibitem[{{Aikawa} {et~al.}(2012){Aikawa}, {Kamuro}, {Sakon}, {Itoh}, {Terada},
  {Noble}, {Pontoppidan}, {Fraser}, {Tamura}, {Kandori}, {Kawamura}, \&
  {Ueno}}]{2012A&A...538A..57A}
{Aikawa}, Y., {Kamuro}, D., {Sakon}, I., {et~al.} 2012, \aap, 538, A57,
  \dodoi{10.1051/0004-6361/201015999}

\bibitem[{{Allamandola} {et~al.}(2021){Allamandola}, {Boersma}, {Lee},
  {Bregman}, \& {Temi}}]{2021ApJ...917L..35A}
{Allamandola}, L.~J., {Boersma}, C., {Lee}, T.~J., {Bregman}, J.~D., \& {Temi},
  P. 2021, \apjl, 917, L35, \dodoi{10.3847/2041-8213/ac17f0}

\bibitem[{{Baba} {et~al.}(2019){Baba}, {Nakagawa}, {Usui}, {Yamagishi}, \&
  {Onaka}}]{2019PASJ...71....2B}
{Baba}, S., {Nakagawa}, T., {Usui}, F., {Yamagishi}, M., \& {Onaka}, T. 2019,
  \pasj, 71, 2, \dodoi{10.1093/pasj/psy131}

\bibitem[{{Bennett} {et~al.}(2010){Bennett}, {Jones}, {Knox}, {Perry}, {Kim},
  \& {Kaiser}}]{2010ApJ...723..641B}
{Bennett}, C.~J., {Jones}, B., {Knox}, E., {et~al.} 2010, \apj, 723, 641,
  \dodoi{10.1088/0004-637X/723/1/641}

\bibitem[{{Bernstein} {et~al.}(1999){Bernstein}, {Sandford}, {Allamandola},
  {Gillette}, {Clemett}, \& {Zare}}]{1999Sci...283.1135B}
{Bernstein}, M.~P., {Sandford}, S.~A., {Allamandola}, L.~J., {et~al.} 1999,
  Science, 283, 1135, \dodoi{10.1126/science.283.5405.1135}

\bibitem[{{Boesgaard} \& {Steigman}(1985)}]{1985ARA&A..23..319B}
{Boesgaard}, A.~M., \& {Steigman}, G. 1985, \araa, 23, 319,
  \dodoi{10.1146/annurev.aa.23.090185.001535}

\bibitem[{{Boogert} {et~al.}(2015){Boogert}, {Gerakines}, \&
  {Whittet}}]{2015ARA&A..53..541B}
{Boogert}, A.~C.~A., {Gerakines}, P.~A., \& {Whittet}, D. C.~B. 2015, \araa,
  53, 541, \dodoi{10.1146/annurev-astro-082214-122348}

\bibitem[{{Boogert} {et~al.}(2000){Boogert}, {Ehrenfreund}, {Gerakines},
  {Tielens}, {Whittet}, {Schutte}, {van Dishoeck}, {de Graauw}, {Decin}, \&
  {Prusti}}]{2000A&A...353..349B}
{Boogert}, A.~C.~A., {Ehrenfreund}, P., {Gerakines}, P.~A., {et~al.} 2000,
  \aap, 353, 349.
\newblock \doarXiv{astro-ph/9909477}

\bibitem[{{Boulanger} {et~al.}(2011){Boulanger}, {Onaka}, {Pilleri}, \&
  {Joblin}}]{2011EAS....46..399B}
{Boulanger}, F., {Onaka}, T., {Pilleri}, P., \& {Joblin}, C. 2011, in EAS
  Publications Series, Vol.~46, EAS Publications Series, ed. C.~{Joblin} \&
  A.~G.~G.~M. {Tielens}, 399--405

\bibitem[{{Buragohain} {et~al.}(2020){Buragohain}, {Pathak}, {Sakon}, \&
  {Onaka}}]{2020ApJ...892...11B}
{Buragohain}, M., {Pathak}, A., {Sakon}, I., \& {Onaka}, T. 2020, \apj, 892,
  11, \dodoi{10.3847/1538-4357/ab733a}

\bibitem[{{Burkhardt} {et~al.}(2021){Burkhardt}, {Loomis}, {Shingledecker},
  {Lee}, {Remijan}, {McCarthy}, \& {McGuire}}]{2021NatAs...5..181B}
{Burkhardt}, A.~M., {Loomis}, R.~A., {Shingledecker}, C.~N., {et~al.} 2021,
  NatAs, 5, 181, \dodoi{10.1038/s41550-020-01253-4}

\bibitem[{{Charnley} {et~al.}(1997){Charnley}, {Tielens}, \&
  {Rodgers}}]{1997ApJ...482L.203C}
{Charnley}, S.~B., {Tielens}, A.~G.~G.~M., \& {Rodgers}, S.~D. 1997, \apjl,
  482, L203, \dodoi{10.1086/310697}

\bibitem[{{Chiar} {et~al.}(2002){Chiar}, {Adamson}, {Pendleton}, {Whittet},
  {Caldwell}, \& {Gibb}}]{2002ApJ...570..198C}
{Chiar}, J.~E., {Adamson}, A.~J., {Pendleton}, Y.~J., {et~al.} 2002, \apj, 570,
  198, \dodoi{10.1086/339570}

\bibitem[{{Cooke} {et~al.}(2018){Cooke}, {Pettini}, \&
  {Steidel}}]{2018ApJ...855..102C}
{Cooke}, R.~J., {Pettini}, M., \& {Steidel}, C.~C. 2018, \apj, 855, 102,
  \dodoi{10.3847/1538-4357/aaab53}

\bibitem[{{Decleir} {et~al.}(2022){Decleir}, {Gordon}, {Andrews}, {Clayton},
  {Cushing}, {Misselt}, {Pendleton}, {Rayner}, {Vacca}, \&
  {Whittet}}]{2022ApJ...930...15D}
{Decleir}, M., {Gordon}, K.~D., {Andrews}, J.~E., {et~al.} 2022, \apj, 930, 15,
  \dodoi{10.3847/1538-4357/ac5dbe}

\bibitem[{{Demyk} {et~al.}(1998){Demyk}, {Dartois}, {D'Hendecourt}, {Jourdain
  de Muizon}, {Heras}, \& {Breitfellner}}]{1998A&A...339..553D}
{Demyk}, K., {Dartois}, E., {D'Hendecourt}, L., {et~al.} 1998, \aap, 339, 553

\bibitem[{{D'Hendecourt} \& {Allamandola}(1986)}]{1986A&AS...64..453D}
{D'Hendecourt}, L.~B., \& {Allamandola}, L.~J. 1986, \aaps, 64, 453

\bibitem[{{Doney} {et~al.}(2016){Doney}, {Candian}, {Mori}, {Onaka}, \&
  {Tielens}}]{2016A&A...586A..65D}
{Doney}, K.~D., {Candian}, A., {Mori}, T., {Onaka}, T., \& {Tielens},
  A.~G.~G.~M. 2016, \aap, 586, A65, \dodoi{10.1051/0004-6361/201526809}

\bibitem[{{Draine}(2006)}]{2006ASPC..348...58D}
{Draine}, B.~T. 2006, in Astronomical Society of the Pacific Conference Series,
  Vol. 348, Astrophysics in the Far Ultraviolet: Five Years of Discovery with
  FUSE, ed. G.~{Sonneborn}, H.~W. {Moos}, \& B.~G. {Andersson}, (San Francisco, CA: ASP), 58

\bibitem[{{Ehrenfreund} {et~al.}(1997){Ehrenfreund}, {Boogert}, {Gerakines},
  {Tielens}, \& {van Dishoeck}}]{1997A&A...328..649E}
{Ehrenfreund}, P., {Boogert}, A.~C.~A., {Gerakines}, P.~A., {Tielens},
  A.~G.~G.~M., \& {van Dishoeck}, E.~F. 1997, \aap, 328, 649

\bibitem[{{Endo} {et~al.}(2021){Endo}, {Sakon}, {Onaka}, {Kimura}, {Kimura},
  {Wada}, {Helton}, {Lau}, {Kebukawa}, {Muramatsu}, {Ogawa}, {Ohkouchi},
  {Nakamura}, \& {Kwok}}]{2021ApJ...917..103E}
{Endo}, I., {Sakon}, I., {Onaka}, T., {et~al.} 2021, \apj, 917, 103,
  \dodoi{10.3847/1538-4357/ac0cf1}

\bibitem[{{Fedoseev} {et~al.}(2016){Fedoseev}, {Chuang}, {van Dishoeck},
  {Ioppolo}, \& {Linnartz}}]{2016MNRAS.460.4297F}
{Fedoseev}, G., {Chuang}, K.~J., {van Dishoeck}, E.~F., {Ioppolo}, S., \&
  {Linnartz}, H. 2016, \mnras, 460, 4297, \dodoi{10.1093/mnras/stw1028}

\bibitem[{{Fedoseev} {et~al.}(2015){Fedoseev}, {Ioppolo}, {Zhao}, {Lamberts},
  \& {Linnartz}}]{2015MNRAS.446..439F}
{Fedoseev}, G., {Ioppolo}, S., {Zhao}, D., {Lamberts}, T., \& {Linnartz}, H.
  2015, \mnras, 446, 439, \dodoi{10.1093/mnras/stu2028}

\bibitem[{{Fraser} {et~al.}(2005){Fraser}, {Bisschop}, {Pontoppidan},
  {Tielens}, \& {van Dishoeck}}]{2005MNRAS.356.1283F}
{Fraser}, H.~J., {Bisschop}, S.~E., {Pontoppidan}, K.~M., {Tielens}, A.
  G.~G.~M., \& {van Dishoeck}, E.~F. 2005, \mnras, 356, 1283,
  \dodoi{10.1111/j.1365-2966.2004.08541.x}

\bibitem[{{Geballe} {et~al.}(1989){Geballe}, {Tielens}, {Allamandola},
  {Moorhouse}, \& {Brand}}]{1989ApJ...341..278G}
{Geballe}, T.~R., {Tielens}, A.~G.~G.~M., {Allamandola}, L.~J., {Moorhouse},
  A., \& {Brand}, P.~W.~J.~L. 1989, \apj, 341, 278, \dodoi{10.1086/167491}

\bibitem[{{Gerakines} {et~al.}(1995){Gerakines}, {Schutte}, {Greenberg}, \&
  {van Dishoeck}}]{1995A&A...296..810G}
{Gerakines}, P.~A., {Schutte}, W.~A., {Greenberg}, J.~M., \& {van Dishoeck},
  E.~F. 1995, \aap, 296, 810.
\newblock \doarXiv{astro-ph/9409076}

\bibitem[{{Giard} {et~al.}(1988){Giard}, {Pajot}, {Lamarre}, {Serra}, {Caux},
  {Gispert}, {Leger}, \& {Rouan}}]{1988A&A...201L...1G}
{Giard}, M., {Pajot}, F., {Lamarre}, J.~M., {et~al.} 1988, \aap, 201, L1

\bibitem[{{Gibb} {et~al.}(2004){Gibb}, {Whittet}, {Boogert}, \&
  {Tielens}}]{2004ApJS..151...35G}
{Gibb}, E.~L., {Whittet}, D.~C.~B., {Boogert}, A.~C.~A., \& {Tielens},
  A.~G.~G.~M. 2004, \apjs, 151, 35, \dodoi{10.1086/381182}

\bibitem[{{Gibb} {et~al.}(2000){Gibb}, {Whittet}, {Schutte}, {Boogert},
  {Chiar}, {Ehrenfreund}, {Gerakines}, {Keane}, {Tielens}, {van Dishoeck}, \&
  {Kerkhof}}]{2000ApJ...536..347G}
{Gibb}, E.~L., {Whittet}, D.~C.~B., {Schutte}, W.~A., {et~al.} 2000, \apj, 536,
  347, \dodoi{10.1086/308940}

\bibitem[{{Gordon} {et~al.}(2021){Gordon}, {Misselt}, {Bouwman}, {Clayton},
  {Decleir}, {Hines}, {Pendleton}, {Rieke}, {Smith}, \&
  {Whittet}}]{2021ApJ...916...33G}
{Gordon}, K.~D., {Misselt}, K.~A., {Bouwman}, J., {et~al.} 2021, \apj, 916, 33,
  \dodoi{10.3847/1538-4357/ac00b7}

\bibitem[{{Hudgins} {et~al.}(2004){Hudgins}, {Bauschlicher}, \&
  {Sandford}}]{2004ApJ...614..770H}
{Hudgins}, D.~M., {Bauschlicher}, C.~W., J., \& {Sandford}, S.~A. 2004, \apj,
  614, 770, \dodoi{10.1086/423930}

\bibitem[{{Hudson} {et~al.}(2001){Hudson}, {Moore}, \&
  {Gerakines}}]{2001ApJ...550.1140H}
{Hudson}, R.~L., {Moore}, M.~H., \& {Gerakines}, P.~A. 2001, \apj, 550, 1140,
  \dodoi{10.1086/319799}

\bibitem[{{IRSA}(2022)}]{https://doi.org/10.26131/irsa543}
{IRSA}. 2022, Spitzer Heritage Archive,  IPAC, \dodoi{10.26131/IRSA543}.
\newblock
  \url{https://catcopy.ipac.caltech.edu/dois/doi.php?id=10.26131/IRSA543}

\bibitem[{{Ishihara} {et~al.}(2010){Ishihara}, {Onaka}, {Kataza}, {Salama},
  {Alfageme}, {Cassatella}, {Cox}, {Garc{\'\i}a-Lario}, {Stephenson}, {Cohen},
  {Fujishiro}, {Fujiwara}, {Hasegawa}, {Ita}, {Kim}, {Matsuhara}, {Murakami},
  {M{\"u}ller}, {Nakagawa}, {Ohyama}, {Oyabu}, {Pyo}, {Sakon}, {Shibai},
  {Takita}, {Tanab{\'e}}, {Uemizu}, {Ueno}, {Usui}, {Wada}, {Watarai},
  {Yamamura}, \& {Yamauchi}}]{2010A&A...514A...1I}
{Ishihara}, D., {Onaka}, T., {Kataza}, H., {et~al.} 2010, \aap, 514, A1,
  \dodoi{10.1051/0004-6361/200913811}

\bibitem[{{Jang} {et~al.}(2022){Jang}, {An}, {Sellgren}, {Ram{\'\i}rez},
  {Boogert}, \& {Schultheis}}]{2022ApJ...930...16J}
{Jang}, D., {An}, D., {Sellgren}, K., {et~al.} 2022, \apj, 930, 16,
  \dodoi{10.3847/1538-4357/ac5d51}

\bibitem[{{Kim} {et~al.}(2022){Kim}, {Lee}, {Jeong}, {Kim}, {Aikawa}, {Noble},
  {Choi}, {Lee}, {Dunham}, {Kim}, \& {Koo}}]{2022ApJ...935..137K}
{Kim}, J., {Lee}, J.-E., {Jeong}, W.-S., {et~al.} 2022, \apj, 935, 137,
  \dodoi{10.3847/1538-4357/ac7f9f}

\bibitem[{{Kim} \& {Koo}(2001)}]{2001ApJ...549..979K}
{Kim}, K.-T., \& {Koo}, B.-C. 2001, \apj, 549, 979, \dodoi{10.1086/319447}

\bibitem[{{Lacy} {et~al.}(1984){Lacy}, {Baas}, {Allamandola}, {Persson},
  {McGregor}, {Lonsdale}, {Geballe}, \& {van de Bult}}]{1984ApJ...276..533L}
{Lacy}, J.~H., {Baas}, F., {Allamandola}, L.~J., {et~al.} 1984, \apj, 276, 533,
  \dodoi{10.1086/161642}

\bibitem[{{Li}(2020)}]{2020NatAs...4..339L}
{Li}, A. 2020, NatAs, 4, 339, \dodoi{10.1038/s41550-020-1051-1}

\bibitem[{{Linsky} {et~al.}(2006){Linsky}, {Draine}, {Moos}, {Jenkins}, {Wood},
  {Oliveira}, {Blair}, {Friedman}, {Gry}, {Knauth}, {Kruk}, {Lacour}, {Lehner},
  {Redfield}, {Shull}, {Sonneborn}, \& {Williger}}]{2006ApJ...647.1106L}
{Linsky}, J.~L., {Draine}, B.~T., {Moos}, H.~W., {et~al.} 2006, \apj, 647,
  1106, \dodoi{10.1086/505556}

\bibitem[{{Mastrapa} {et~al.}(2009){Mastrapa}, {Sandford}, {Roush},
  {Cruikshank}, \& {Dalle Ore}}]{2009ApJ...701.1347M}
{Mastrapa}, R.~M., {Sandford}, S.~A., {Roush}, T.~L., {Cruikshank}, D.~P., \&
  {Dalle Ore}, C.~M. 2009, \apj, 701, 1347,
  \dodoi{10.1088/0004-637X/701/2/1347}

\bibitem[{{McGuire} {et~al.}(2018){McGuire}, {Burkhardt}, {Kalenskii},
  {Shingledecker}, {Remijan}, {Herbst}, \& {McCarthy}}]{2018Sci...359..202M}
{McGuire}, B.~A., {Burkhardt}, A.~M., {Kalenskii}, S., {et~al.} 2018, Science,
  359, 202, \dodoi{10.1126/science.aao4890}

\bibitem[{{McGuire} {et~al.}(2021){McGuire}, {Loomis}, {Burkhardt}, {Lee},
  {Shingledecker}, {Charnley}, {Cooke}, {Cordiner}, {Herbst}, {Kalenskii},
  {Siebert}, {Willis}, {Xue}, {Remijan}, \& {McCarthy}}]{2021Sci...371.1265M}
{McGuire}, B.~A., {Loomis}, R.~A., {Burkhardt}, A.~M., {et~al.} 2021, Science,
  371, 1265, \dodoi{10.1126/science.abb7535}

\bibitem[{{Molinari} {et~al.}(2016){Molinari}, {Schisano}, {Elia},
  {Pestalozzi}, {Traficante}, {Pezzuto}, {Swinyard}, {Noriega-Crespo}, {Bally},
  {Moore}, {Plume}, {Zavagno}, {di Giorgio}, {Liu}, {Pilbratt}, {Mottram},
  {Russeil}, {Piazzo}, {Veneziani}, {Benedettini}, {Calzoletti}, {Faustini},
  {Natoli}, {Piacentini}, {Merello}, {Palmese}, {Del Grande}, {Polychroni},
  {Rygl}, {Polenta}, {Barlow}, {Bernard}, {Martin}, {Testi}, {Ali},
  {Andr{\'e}}, {Beltr{\'a}n}, {Billot}, {Carey}, {Cesaroni}, {Compi{\`e}gne},
  {Eden}, {Fukui}, {Garcia-Lario}, {Hoare}, {Huang}, {Joncas}, {Lim}, {Lord},
  {Martinavarro-Armengol}, {Motte}, {Paladini}, {Paradis}, {Peretto},
  {Robitaille}, {Schilke}, {Schneider}, {Schulz}, {Sibthorpe}, {Strafella},
  {Thompson}, {Umana}, {Ward-Thompson}, \& {Wyrowski}}]{2016A&A...591A.149M}
{Molinari}, S., {Schisano}, E., {Elia}, D., {et~al.} 2016, \aap, 591, A149,
  \dodoi{10.1051/0004-6361/201526380}

\bibitem[{{Moneti} {et~al.}(1994){Moneti}, {Glass}, \&
  {Moorwood}}]{1994MNRAS.268..194M}
{Moneti}, A., {Glass}, I.~S., \& {Moorwood}, A.~F.~M. 1994, \mnras, 268, 194,
  \dodoi{10.1093/mnras/268.1.194}

\bibitem[{{Mori} {et~al.}(2022){Mori}, {Onaka}, {Sakon}, {Buragohain},
  {Takahata}, {Sano}, \& {Pathak}}]{2022ApJ...933...35M}
{Mori}, T., {Onaka}, T., {Sakon}, I., {et~al.} 2022, \apj, 933, 35,
  \dodoi{10.3847/1538-4357/ac71ae}

\bibitem[{{Mori} {et~al.}(2014){Mori}, {Onaka}, {Sakon}, {Ishihara},
  {Shimonishi}, {Ohsawa}, \& {Bell}}]{2014ApJ...784...53M}
{Mori}, T.~I., {Onaka}, T., {Sakon}, I., {et~al.} 2014, \apj, 784, 53,
  \dodoi{10.1088/0004-637X/784/1/53}

\bibitem[{{Moultaka} {et~al.}(2004){Moultaka}, {Eckart}, {Viehmann}, {Mouawad},
  {Straubmeier}, {Ott}, \& {Sch{\"o}del}}]{2004A&A...425..529M}
{Moultaka}, J., {Eckart}, A., {Viehmann}, T., {et~al.} 2004, \aap, 425, 529,
  \dodoi{10.1051/0004-6361:20035807}

\bibitem[{Nakamura {et~al.}(2022)Nakamura, Kobayashi, Tanaka, KUNIHIRO,
  KITAGAWA, POTISZIL, OTA, SAKAGUCHI, YAMANAKA, RATNAYAKE, TRIPATHI, KUMAR,
  AVRAMESCU, TSUCHIDA, YACHI, MIURA, ABE, FUKAI, FURUYA, HATAKEDA, HAYASHI,
  HITOMI, KUMAGAI, MIYAZAKI, NAKATO, NISHIMURA, OKADA, SOEJIMA, SUGITA, SUZUKI,
  USUI, YADA, YAMAMOTO, YOGATA, YOSHITAKE, ARAKAWA, FUJII, HAYAKAWA, HIRATA,
  HIRATA, HONDA, HONDA, HOSODA, ichi IIJIMA, IKEDA, ISHIGURO, ISHIHARA, IWATA,
  KAWAHARA, KIKUCHI, KITAZATO, MATSUMOTO, MATSUOKA, MICHIKAMI, MIMASU, MIURA,
  MOROTA, NAKAZAWA, NAMIKI, NODA, NOGUCHI, OGAWA, OGAWA, OKAMOTO, ONO, OZAKI,
  SAIKI, SAKATANI, SAWADA, SENSHU, SHIMAKI, SHIRAI, TAKEI, TAKEUCHI, TANAKA,
  TATSUMI, TERUI, TSUKIZAKI, WADA, YAMADA, YAMADA, YAMAMOTO, YANO, YOKOTA,
  YOSHIHARA, YOSHIKAWA, YOSHIKAWA, FUJIMOTO, ichiro WATANABE, \&
  TSUDA}]{EizoNAKAMURA2022PJA9806B-01}
Nakamura, E., Kobayashi, K., Tanaka, R., {et~al.} 2022, PJAB, 
  98, 227, \dodoi{10.2183/pjab.98.015}

\bibitem[{{Neugebauer} {et~al.}(2019){Neugebauer}, {Habing}, {van Duinen},
  {Aumann}, {Baud}, {Beichman}, {Beintema}, {Boggess}, {Clegg}, {de Jong},
  {Emerson}, {Gautier}, {Gillett}, {Harris}, {Hauser}, {Houck}, {Jennings},
  {Low}, {Marsden}, {Miley}, {Olnon}, {Pottasch}, {Raimond}, {Rowan-Robinson},
  {Soifer}, {Walker}, {Wesselius}, \& {Young}}]{https://doi.org/10.26131/irsa4}
{Neugebauer}, G., {Habing}, H.~J., {van Duinen}, R., {et~al.} 2019, IRAS Point
  Source Catalog v2.1 (PSC),  IPAC, \dodoi{10.26131/IRSA4}.

\bibitem[{{Noble} {et~al.}(2013){Noble}, {Fraser}, {Aikawa}, {Pontoppidan}, \&
  {Sakon}}]{2013ApJ...775...85N}
{Noble}, J.~A., {Fraser}, H.~J., {Aikawa}, Y., {Pontoppidan}, K.~M., \&
  {Sakon}, I. 2013, \apj, 775, 85, \dodoi{10.1088/0004-637X/775/2/85}

\bibitem[{{{\"O}berg} {et~al.}(2011){{\"O}berg}, {Boogert}, {Pontoppidan}, {van
  den Broek}, {van Dishoeck}, {Bottinelli}, {Blake}, \&
  {Evans}}]{2011ApJ...740..109O}
{{\"O}berg}, K.~I., {Boogert}, A.~C.~A., {Pontoppidan}, K.~M., {et~al.} 2011,
  \apj, 740, 109, \dodoi{10.1088/0004-637X/740/2/109}

\bibitem[{{Ohyama} {et~al.}(2007){Ohyama}, {Onaka}, {Matsuhara}, {Wada}, {Kim},
  {Fujishiro}, {Uemizu}, {Sakon}, {Cohen}, {Ishigaki}, {Ishihara}, {Ita},
  {Kataza}, {Matsumoto}, {Murakami}, {Oyabu}, {Tanab{\'e}}, {Takagi}, {Ueno},
  {Usui}, {Watarai}, {Pearson}, {Takeyama}, {Yamamuro}, \&
  {Ikeda}}]{2007PASJ...59S.411O}
{Ohyama}, Y., {Onaka}, T., {Matsuhara}, H., {et~al.} 2007, \pasj, 59, S411,
  \dodoi{10.1093/pasj/59.sp2.S411}

\bibitem[{{Onaka} {et~al.}(2021){Onaka}, {Kimura}, {Sakon}, \&
  {Shimonishi}}]{2021ApJ...916...75O}
{Onaka}, T., {Kimura}, T., {Sakon}, I., \& {Shimonishi}, T. 2021, \apj, 916,
  75, \dodoi{10.3847/1538-4357/ac0531}

\bibitem[{{Onaka} {et~al.}(2014){Onaka}, {Mori}, {Sakon}, {Ohsawa}, {Kaneda},
  {Okada}, \& {Tanaka}}]{2014ApJ...780..114O}
{Onaka}, T., {Mori}, T.~I., {Sakon}, I., {et~al.} 2014, \apj, 780, 114,
  \dodoi{10.1088/0004-637X/780/2/114}

\bibitem[{{Onaka} {et~al.}(2007){Onaka}, {Matsuhara}, {Wada}, {Fujishiro},
  {Fujiwara}, {Ishigaki}, {Ishihara}, {Ita}, {Kataza}, {Kim}, {Matsumoto},
  {Murakami}, {Ohyama}, {Oyabu}, {Sakon}, {Tanab{\'e}}, {Takagi}, {Uemizu},
  {Ueno}, {Usui}, {Watarai}, {Cohen}, {Enya}, {Ootsubo}, {Pearson}, {Takeyama},
  {Yamamuro}, \& {Ikeda}}]{2007PASJ...59S.401O}
{Onaka}, T., {Matsuhara}, H., {Wada}, T., {et~al.} 2007, \pasj, 59, S401,
  \dodoi{10.1093/pasj/59.sp2.S401}

\bibitem[{{Onaka} {et~al.}(2010){Onaka}, {Matsuhara}, {Wada}, {Ishihara},
  {Ita}, {Ohyama}, {Ootsubo}, {Oyabu}, {Sakon}, {Shimonishi}, {Takita},
  {Tanab{\`e}}, {Usui}, \& {Murakami}}]{2010SPIE.7731E..0MO}
{Onaka}, T., {Matsuhara}, H., {Wada}, T., {et~al.} 2010, in Society of
  Photo-Optical Instrumentation Engineers (SPIE) Conference Series, Vol. 7731,
  Space Telescopes and Instrumentation 2010: Optical, Infrared, and Millimeter
  Wave, ed. J.~{Oschmann}, Jacobus~M., M.~C. {Clampin}, \& H.~A. {MacEwen},
  77310M

\bibitem[{{Ossenkopf} {et~al.}(1992){Ossenkopf}, {Henning}, \&
  {Mathis}}]{1992A&A...261..567O}
{Ossenkopf}, V., {Henning}, T., \& {Mathis}, J.~S. 1992, \aap, 261, 567

\bibitem[{{Peeters} {et~al.}(2004){Peeters}, {Allamandola}, {Bauschlicher},
  {Hudgins}, {Sandford}, \& {Tielens}}]{2004ApJ...604..252P}
{Peeters}, E., {Allamandola}, L.~J., {Bauschlicher}, C.~W., J., {et~al.} 2004,
  \apj, 604, 252, \dodoi{10.1086/381866}

\bibitem[{{Pendleton} {et~al.}(1999){Pendleton}, {Tielens}, {Tokunaga}, \&
  {Bernstein}}]{1999ApJ...513..294P}
{Pendleton}, Y.~J., {Tielens}, A.~G.~G.~M., {Tokunaga}, A.~T., \& {Bernstein},
  M.~P. 1999, \apj, 513, 294, \dodoi{10.1086/306827}

\bibitem[{{Pontoppidan} {et~al.}(2003){Pontoppidan}, {Fraser}, {Dartois},
  {Thi}, {van Dishoeck}, {Boogert}, {d'Hendecourt}, {Tielens}, \&
  {Bisschop}}]{2003A&A...408..981P}
{Pontoppidan}, K.~M., {Fraser}, H.~J., {Dartois}, E., {et~al.} 2003, \aap, 408,
  981, \dodoi{10.1051/0004-6361:20031030}

\bibitem[{{Prodanovi{\'c}} {et~al.}(2010){Prodanovi{\'c}}, {Steigman}, \&
  {Fields}}]{2010MNRAS.406.1108P}
{Prodanovi{\'c}}, T., {Steigman}, G., \& {Fields}, B.~D. 2010, \mnras, 406,
  1108, \dodoi{10.1111/j.1365-2966.2010.16734.x}

\bibitem[{{Ram{\'\i}rez} {et~al.}(2021){Ram{\'\i}rez}, {Arendt}, {Sellgren},
  {Stolovy}, {Cotera}, {Smith}, \&
  {Yusef-Zadeh}}]{https://doi.org/10.26131/irsa505}
---. 2021, Point Sources from a Spitzer/IRAC Survey of the Galactic Center,
  IPAC, \dodoi{10.26131/IRSA505}.

\bibitem[{{Ram{\'\i}rez} {et~al.}(2008){Ram{\'\i}rez}, {Arendt}, {Sellgren},
  {Stolovy}, {Cotera}, {Smith}, \& {Yusef-Zadeh}}]{2008ApJS..175..147R}
{Ram{\'\i}rez}, S.~V., {Arendt}, R.~G., {Sellgren}, K., {et~al.} 2008, \apjs,
  175, 147, \dodoi{10.1086/524015}

\bibitem[{{Raunier} {et~al.}(2003{\natexlab{a}}){Raunier}, {Chiavassa},
  {Allouche}, {Marinelli}, \& {Aycard}}]{2003CP....288..197R}
{Raunier}, S., {Chiavassa}, T., {Allouche}, A., {Marinelli}, F., \& {Aycard},
  J.-P. 2003{\natexlab{a}}, Chemical Physics, 288, 197,
  \dodoi{10.1016/S0301-0104(03)00024-7}

\bibitem[{{Raunier} {et~al.}(2003{\natexlab{b}}){Raunier}, {Chiavassa},
  {Marinelli}, {Allouche}, \& {Aycard}}]{2003CPL...368..594R}
{Raunier}, S., {Chiavassa}, T., {Marinelli}, F., {Allouche}, A., \& {Aycard},
  J.~P. 2003{\natexlab{b}}, Chemical Physics Letters, 368, 594,
  \dodoi{10.1016/S0009-2614(02)01919-X}

\bibitem[{{Ricca} {et~al.}(2021){Ricca}, {Boersma}, \&
  {Peeters}}]{2021ApJ...923..202R}
{Ricca}, A., {Boersma}, C., \& {Peeters}, E. 2021, \apj, 923, 202,
  \dodoi{10.3847/1538-4357/ac28fc}

\bibitem[{{Romano} {et~al.}(2006){Romano}, {Tosi}, {Chiappini}, \&
  {Matteucci}}]{2006MNRAS.369..295R}
{Romano}, D., {Tosi}, M., {Chiappini}, C., \& {Matteucci}, F. 2006, \mnras,
  369, 295, \dodoi{10.1111/j.1365-2966.2006.10287.x}

\bibitem[{{Sandford} {et~al.}(2000){Sandford}, {Bernstein}, {Allamandola},
  {Gillette}, \& {Zare}}]{2000ApJ...538..691S}
{Sandford}, S.~A., {Bernstein}, M.~P., {Allamandola}, L.~J., {Gillette}, J.~S.,
  \& {Zare}, R.~N. 2000, \apj, 538, 691, \dodoi{10.1086/309147}

\bibitem[{{Sandford} {et~al.}(2001){Sandford}, {Bernstein}, \&
  {Dworkin}}]{2001M&PS...36.1117S}
{Sandford}, S.~A., {Bernstein}, M.~P., \& {Dworkin}, J.~P. 2001, \maps, 36,
  1117, \dodoi{10.1111/j.1945-5100.2001.tb01948.x}

\bibitem[{{Shimonishi} {et~al.}(2010){Shimonishi}, {Onaka}, {Kato}, {Sakon},
  {Ita}, {Kawamura}, \& {Kaneda}}]{2010A&A...514A..12S}
{Shimonishi}, T., {Onaka}, T., {Kato}, D., {et~al.} 2010, \aap, 514, A12,
  \dodoi{10.1051/0004-6361/200913815}

\bibitem[{{Skrutskie} {et~al.}(2003){Skrutskie}, {Cutri}, {Stiening},
  {Weinberg}, {Schneider}, {Carpenter}, {Beichman}, {Capps}, {Chester},
  {Elias}, {Huchra}, {Liebert}, {Lonsdale}, {Monet}, {Price}, {Seitzer},
  {Jarrett}, {Kirkpatrick}, {Gizis}, {Howard}, {Evans}, {Fowler}, {Fullmer},
  {Hurt}, {Light}, {Kopan}, {Marsh}, {McCallon}, {Tam}, {Van Dyk}, \&
  {Wheelock}}]{https://doi.org/10.26131/irsa2}
{Skrutskie}, M.~F., {Cutri}, R.~M., {Stiening}, R., {et~al.} 2003, 2MASS
  All-Sky Point Source Catalog,  IPAC, \dodoi{10.26131/IRSA2}.

\bibitem[{{Skrutskie} {et~al.}(2006){Skrutskie}, {Cutri}, {Stiening},
  {Weinberg}, {Schneider}, {Carpenter}, {Beichman}, {Capps}, {Chester},
  {Elias}, {Huchra}, {Liebert}, {Lonsdale}, {Monet}, {Price}, {Seitzer},
  {Jarrett}, {Kirkpatrick}, {Gizis}, {Howard}, {Evans}, {Fowler}, {Fullmer},
  {Hurt}, {Light}, {Kopan}, {Marsh}, {McCallon}, {Tam}, {Van Dyk}, \&
  {Wheelock}}]{2006AJ....131.1163S}
---. 2006, \aj, 131, 1163, \dodoi{10.1086/498708}

\bibitem[{{Snyder} \& {Buhl}(1972)}]{1972ApJ...177..619S}
{Snyder}, L.~E., \& {Buhl}, D. 1972, \apj, 177, 619, \dodoi{10.1086/151739}

\bibitem[{{Spoon} {et~al.}(2003){Spoon}, {Moorwood}, {Pontoppidan}, {Cami},
  {Kregel}, {Lutz}, \& {Tielens}}]{2003A&A...402..499S}
{Spoon}, H.~W.~W., {Moorwood}, A.~F.~M., {Pontoppidan}, K.~M., {et~al.} 2003,
  \aap, 402, 499, \dodoi{10.1051/0004-6361:20030290}

\bibitem[{{Storey} \& {Hummer}(1995)}]{1995MNRAS.272...41S}
{Storey}, P.~J., \& {Hummer}, D.~G. 1995, \mnras, 272, 41,
  \dodoi{10.1093/mnras/272.1.41}

\bibitem[{{Taban} {et~al.}(2003){Taban}, {Schutte}, {Pontoppidan}, \& {van
  Dishoeck}}]{2003A&A...399..169T}
{Taban}, I.~M., {Schutte}, W.~A., {Pontoppidan}, K.~M., \& {van Dishoeck},
  E.~F. 2003, \aap, 399, 169, \dodoi{10.1051/0004-6361:20021798}

\bibitem[{{Tanaka} {et~al.}(1996){Tanaka}, {Matsumoto}, {Murakami}, {Kawada},
  {Noda}, \& {Matsuura}}]{1996PASJ...48L..53T}
{Tanaka}, M., {Matsumoto}, T., {Murakami}, H., {et~al.} 1996, \pasj, 48, L53,
  \dodoi{10.1093/pasj/48.5.L53}

\bibitem[{{Tielens}(1983)}]{1983A&A...119..177T}
{Tielens}, A.~G.~G.~M. 1983, \aap, 119, 177

\bibitem[{{Tielens}(1997)}]{1997AIPC..402..523T}
{Tielens}, A.~G.~G.~M. 1997, in American Institute of Physics Conference
  Series, Vol. 402, Astrophysical implications of the laboratory study of
  presolar materials, ed. T.~J. {Bernatowicz} \& E.~{Zinner}, 523--544

\bibitem[{{Tielens}(2008)}]{2008ARA&A..46..289T}
---. 2008, \araa, 46, 289, \dodoi{10.1146/annurev.astro.46.060407.145211}

\bibitem[{{Tosi}(2010)}]{2010IAUS..268..153T}
{Tosi}, M. 2010, in Proc. IAU Symp. 268, Light Elements in the Universe, ed. C.~{Charbonnel},
  M.~{Tosi}, F.~{Primas}, \& C.~{Chiappini}, (Cambridge: Cambridge Univ. Press), 153

\bibitem[{{Tsumura} {et~al.}(2013){Tsumura}, {Matsumoto}, {Matsuura}, {Sakon},
  {Tanaka}, \& {Wada}}]{2013PASJ...65..120T}
{Tsumura}, K., {Matsumoto}, T., {Matsuura}, S., {et~al.} 2013, \pasj, 65, 120,
  \dodoi{10.1093/pasj/65.6.120}

\bibitem[{{van Broekhuizen} {et~al.}(2004){van Broekhuizen}, {Keane}, \&
  {Schutte}}]{2004A&A...415..425V}
{van Broekhuizen}, F.~A., {Keane}, J.~V., \& {Schutte}, W.~A. 2004, \aap, 415,
  425, \dodoi{10.1051/0004-6361:20034161}

\bibitem[{{van Broekhuizen} {et~al.}(2005){van Broekhuizen}, {Pontoppidan},
  {Fraser}, \& {van Dishoeck}}]{2005A&A...441..249V}
{van Broekhuizen}, F.~A., {Pontoppidan}, K.~M., {Fraser}, H.~J., \& {van
  Dishoeck}, E.~F. 2005, \aap, 441, 249, \dodoi{10.1051/0004-6361:20041711}

\bibitem[{{Vats} {et~al.}(2022){Vats}, {Pathak}, {Onaka}, {Buragohain},
  {Sakon}, \& {Endo}}]{2022PASJ...74..161V}
{Vats}, A., {Pathak}, A., {Onaka}, T., {et~al.} 2022, \pasj, 74, 161,
  \dodoi{10.1093/pasj/psab116}

\bibitem[{{Watanabe} \& {Kouchi}(2002)}]{2002ApJ...567..651W}
{Watanabe}, N., \& {Kouchi}, A. 2002, \apj, 567, 651, \dodoi{10.1086/338491}

\bibitem[{{Wiersma} {et~al.}(2020){Wiersma}, {Candian}, {Bakker}, {Martens},
  {Berden}, {Oomens}, {Buma}, \& {Petrignani}}]{2020A&A...635A...9W}
{Wiersma}, S.~D., {Candian}, A., {Bakker}, J.~M., {et~al.} 2020, \aap, 635, A9,
  \dodoi{10.1051/0004-6361/201936982}

\bibitem[{{Wright} {et~al.}(2010){Wright}, {Eisenhardt}, {Mainzer}, {Ressler},
  {Cutri}, {Jarrett}, {Kirkpatrick}, {Padgett}, {McMillan}, {Skrutskie},
  {Stanford}, {Cohen}, {Walker}, {Mather}, {Leisawitz}, {Gautier}, {McLean},
  {Benford}, {Lonsdale}, {Blain}, {Mendez}, {Irace}, {Duval}, {Liu}, {Royer},
  {Heinrichsen}, {Howard}, {Shannon}, {Kendall}, {Walsh}, {Larsen}, {Cardon},
  {Schick}, {Schwalm}, {Abid}, {Fabinsky}, {Naes}, \&
  {Tsai}}]{2010AJ....140.1868W}
{Wright}, E.~L., {Eisenhardt}, P. R.~M., {Mainzer}, A.~K., {et~al.} 2010, \aj,
  140, 1868, \dodoi{10.1088/0004-6256/140/6/1868}

\bibitem[{{Wright} {et~al.}(2019){Wright}, {Eisenhardt}, {Mainzer}, {Ressler},
  {Cutri}, {Jarrett}, {Kirkpatrick}, {Padgett}, {McMillan}, {Skrutskie},
  {Stanford}, {Cohen}, {Walker}, {Mather}, {Leisawitz}, {Gautier}, {McLean},
  {Benford}, {Lonsdale}, {Blain}, {Mendez}, {Irace}, {Duval}, {Liu}, {Royer},
  {Heinrichsen}, {Howard}, {Shannon}, {Kendall}, {Walsh}, {Larsen}, {Cardon},
  {Schick}, {Schwalm}, {Abid}, {Fabinsky}, {Naes}, \&
  {Tsai}}]{https://doi.org/10.26131/irsa1}
---. 2019, AllWISE Source Catalog,  IPAC, \dodoi{10.26131/IRSA1}.

\bibitem[{{Yang} {et~al.}(2016){Yang}, {Li}, {Glaser}, \&
  {Zhong}}]{2016ApJ...825...22Y}
{Yang}, X.~J., {Li}, A., {Glaser}, R., \& {Zhong}, J.~X. 2016, \apj, 825, 22,
  \dodoi{10.3847/0004-637X/825/1/22}

\bibitem[{{Yang} {et~al.}(2021){Yang}, {Li}, {He}, \&
  {Glaser}}]{2021ApJS..255...23Y}
{Yang}, X.~J., {Li}, A., {He}, C.~Y., \& {Glaser}, R. 2021, \apjs, 255, 23,
  \dodoi{10.3847/1538-4365/ac0bb5}

\bibitem[{{Zavarygin} {et~al.}(2018){Zavarygin}, {Webb}, {Dumont}, \&
  {Riemer-S{\o}rensen}}]{2018MNRAS.477.5536Z}
{Zavarygin}, E.~O., {Webb}, J.~K., {Dumont}, V., \& {Riemer-S{\o}rensen}, S.
  2018, \mnras, 477, 5536, \dodoi{10.1093/mnras/sty1003}

\end{thebibliography}
\bibliographystyle{aasjournal}



\end{document}